\documentclass[sigplan,10pt]{acmart}
\renewcommand\footnotetextcopyrightpermission[1]{}

\usepackage[english]{babel}
\usepackage{blindtext}

\usepackage{subfigure}

\usepackage{rotating}

\usepackage{array}%
\usepackage{footmisc}
\usepackage{pifont}
\usepackage{soul}
\usepackage{stfloats}
\usepackage{xcolor,colortbl}

\usepackage{booktabs}
\usepackage{array}
\usepackage{blindtext}
\usepackage{array}%
\usepackage{footmisc}
\usepackage{xcolor,colortbl}

\usepackage{verbatim}

\usepackage{multirow}
\usepackage[utf8]{inputenc}

\usepackage{makecell}
\usepackage{cleveref}
\usepackage{setspace}
\usepackage{stfloats}
\usepackage{enumitem}
\usepackage[noend]{algpseudocode}
\usepackage[linesnumbered,lined,commentsnumbered,ruled]{algorithm2e}

\usepackage{xcolor,colortbl}

\usepackage{cleveref}

\renewcommand\footnotetextcopyrightpermission[1]{}

\crefname{section}{§}{§§}

\newcommand{\addData}{\textcolor{black}}
\newcommand{\adddata}{\textcolor{black}}
\newcommand{\tcr}{\textcolor{black}}

\newcommand{\system}{DualDecoder\xspace}

\newcommand{\residency}{KV-cache-auxiliary residency\xspace}
\newcommand{\spectoken}{speculative token\xspace}
\newcommand{\spectokens}{speculative tokens\xspace}

\SetKwInput{KwInput}{Input}     
\SetKw{KwAll}{all\ }
\SetKw{KwAny}{any\ }
\SetKwInput{KwOutput}{Output} 
\SetKwProg{Fn}{Function}{:}{end}

\SetAlFnt{\footnotesize}

\graphicspath{{figures/}}

\acmDOI{}

\acmISBN{}

\acmPrice{}

\begin{document}

\title{DualDecoder: Accelerate Long Context LLM Inference by Predictive Prefetch}

\author{Zuning Liang}

\affiliation{%
  \institution{Fudan University \\Shanghai Innovation Institute}
  \country{China}
}
\email{25113090054@m.fudan.edu.cn}

\author{Zhiyi Yao}
\affiliation{%
  \institution{Fudan University}
  \country{China}
}
\email{23110720054@m.fudan.edu.cn}

\author{Qi Chen}
\affiliation{%
  \institution{Fudan University}
  \country{China}
}
\email{22307130007@m.fudan.edu.cn}

\author{Yuedong Xu}
\affiliation{%
  \institution{Fudan University}
  \country{China}
}
\email{ydxu@fudan.edu.cn}

\author{Hao Dai}
\affiliation{%
  \institution{Ant Group}
  \country{China}
}
\email{dh183333@antgroup.com}

\author{Zhiqiang Ding}
\affiliation{%
  \institution{Ant Group}
  \country{China}
}
\email{chucai.dzq@antgroup.com}

\author{Tongkai Yang}
\affiliation{%
  \institution{Ant Group}
  \country{China}
}
\email{tongkai.ytk@antgroup.com}

\author{Jinlong Hou}
\affiliation{%
  \institution{Fudan University \\ Shanghai Innovation Institute}
  \country{China}
}
\email{houjinlong@sii.edu.cn}
\author{Yuan Cheng}
\affiliation{%
  \institution{Fudan University \\ Shanghai Innovation Institute}
  \country{China}
}
\email{cheng_yuan@fudan.edu.cn}
\begin{abstract}

Long-context inference is becoming a fundamental capability for modern LLM serving, especially driven by emerging agentic applications. Yet it faces a severe memory wall that the KV cache scales proportionally with increasing context length and request concurrency.
Existing sparse KV cache methods offload most KV entries to host memory and retrieve only the critical KV entries needed by each decoding step. However, they commonly introduce substantial auxiliary states in GPU memory for KV retrieval management. Our measurements show that these often-overlooked auxiliary states introduce significant memory overhead and become a new bottleneck under high-concurrency workloads.

In this paper, we present \system, a lightweight serving system for long-context LLM inference that enables efficient sparse KV cache retrieval from host memory. Our key insight is that the critical KV entries required for decoding the next token can be accurately predicted from the preceding speculated token. This predictability enables KV retrieval to be proactively prefetched and overlapped with decoding computation, effectively eliminating the GPU memory overhead of auxiliary states. To achieve this prefetching efficiently, \system leverages a novel dual-token decoding pipeline that accurately identifies critical KV entries with negligible computational overhead, and designs a layer-aware transfer schedule to overlap KV prefetching with model computation and a layer-scoped memory manager to reduce the GPU runtime buffer. Experimental results show that \system improves decoding throughput by up to \addData{2.62$\times$} over state-of-the-art systems while preserving decoding latency and model quality.

\end{abstract}

\settopmatter{printfolios=true}
\maketitle
\pagestyle{plain}

\begin{sloppypar}
\section{Introduction}
\label{sec: introduction}

In large language model (LLM) and agentic AI services, long contexts significantly improve inference accuracy and response coherence, enabling the execution of complex real-world tasks such as full-document understanding~\cite{DBLP:conf/acl/BaiTZ0WLCX0D0L25,hsieh2024rulerwhatsrealcontext, DBLP:conf/icml/DingZZXSX0Y24} and large-scale codebase analysis without context fragmentation~\cite{DBLP:journals/corr/abs-2505-22101,DBLP:conf/emnlp/0019ZHLHDGZ024}. Modern LLMs increasingly support context lengths ranging from hundreds of thousands to even millions of tokens~\cite{DBLP:journals/corr/abs-2512-02556,DBLP:conf/icml/ShangZWZL0C025}. A major challenge in long-context inference is the limited capacity of GPU memory. To avoid repeatedly performing expensive extra matrix multiplications, LLMs cache intermediate key and value (KV) tensors generated during autoregressive decoding steps~\cite{DBLP:journals/corr/abs-1911-02150,DBLP:conf/sigcomm/LiuLCRHZDY0AMHH24}. However, the KV cache grows proportionally with context length and request batch size, and can quickly occupy a substantial portion of the limited GPU memory~\cite{DBLP:conf/osdi/LeeLSS24,kwon2023vllm,DBLP:conf/icml/LiuYJZXBC024}. Consequently, reducing the memory footprint of the KV cache has become a critical problem in enabling efficient long-context inference.

Dynamic sparse KV cache techniques~\cite{zhang2023h2o,DBLP:conf/icml/TangZZXKH24} provide an effective direction for mitigating this memory pressure. Since each decoding step usually attends to only a small subset of historical tokens~\cite{DBLP:conf/nips/LiHYVLYCLC24}, instead of keeping the entire KV cache in GPU memory, these systems offload most KV entries to host memory and retrieve only the selected sparse KV entries needed by the current attention computation. This design substantially reduces the memory footprint of GPU-resident KV entries and becomes a promising approach for high-throughput, long-context LLM serving (\cref{sec: background}).

However, the selected KV entries must be gathered from host memory and transferred to GPU memory before attention consumes them, while CPU-to-GPU bandwidth is much lower than GPU memory bandwidth. To reduce this retrieval overhead, existing systems keep additional auxiliary states in GPU memory. For example, such states may reconstruct selected key entries~\cite{lin2024shadowkv,zhang2023h2o} or guide future KV prefetching~\cite{chen2024specache}. These designs are effective because they reduce the amount of exposed CPU-to-GPU transfer on the decoding critical path. Yet they also introduce a new memory cost that is easy to overlook. We refer to the GPU memory occupied by such auxiliary state as \textit{\residency}. Our measurements show that this residency can dominate the memory usage and even consume \addData{64\%} of GPU memory in heavy workloads. Under high-concurrency workloads, \residency becomes a new bottleneck that limits the maximum request batch size and prevents dynamic sparse KV cache systems from reaching the serving throughput that the hardware could otherwise support (\cref{sec: motivation}).

In this paper, our key observation is that sparse KV retrieval can be predicted across adjacent decoding steps. In dynamic sparse KV cache mechanisms, the retrieval index of a decoding step is computed from the token being processed and a static GPU-resident KV landmarks. Therefore, once a useful prediction of the next token is available, we can estimate the retrieval index of the next step before that step reaches attention. More importantly, our measurements show that even when token prediction is imperfect, it can still provide sufficient information to predict most of the KV entries needed by the following decoding step. This predictability creates a new opportunity for long-context serving. Instead of waiting for the true retrieval index and then fetching selected KV entries on the critical path, the system can issue KV retrieval earlier and overlap host-to-device transfer with decoding computation. In this way, sparse KV retrieval can be conducted fast without relying on large GPU-resident auxiliary states, leaving more GPU memory available for larger request batches (\cref{sec: Predictive KV Retrieval}).

Based on this observation, we present \system, a lightweight long-context LLM serving system that makes sparse KV retrieval predictive rather than residency heavy. The core of \system is to predict the retrieval index of future decoding work and prefetch the corresponding sparse KV entries from host memory before they are consumed by attention. This approach allows \system to overlap KV retrieval with model computation, thereby preserving the latency benefit of dynamic sparse KV cache without maintaining large GPU-resident auxiliary state. Realizing this approach in a practical decoding pipeline requires addressing three challenges. First, the system must obtain retrieval guidance without turning each decoding step into a much heavier model execution. To address this issue, \system introduces a dual-token decoding pipeline that generates the normal output token and the retrieval-guiding token within the same decoding flow (\cref{sec: Dual-Token Decoding Pipeline}). Second, predicted KV entries must be fetched early enough to hide CPU-to-GPU transfer latency, but not too early that they occupy much GPU memory. \system addresses this tension with a layer-aware transfer schedule that efficiently overlaps KV movement with model layer execution and corrects prediction misses on the critical path (\cref{sec: Speculative KV Cache Transfer Schedule}). Third, predictive prefetching must not recreate the memory pressure that it is designed to remove. \system therefore uses a layer-scoped KV memory manager that significantly reduces the runtime sparse KV buffer size and coordinates asynchronous KV transfer with attention computation (\cref{sec: Layer-Scoped KV Memory Management}). 

\noindent\textbf{Contributions.}
Our main contributions are summarized as follows:

\begin{itemize}[leftmargin=0.4cm, itemsep=1pt]

\item We identify \textit{KV retrieval predictability} as a new opportunity for long-context LLM serving. In dynamic sparse KV cache systems, the retrieval index of a future decoding step can be predicted before that step reaches attention. To the best of our knowledge, this is the first work to characterize and exploit KV retrieval predictability (\cref{sec: Predictive KV Retrieval}).

\item We design \system, a lightweight predictive retrieval system that replaces residency-heavy sparse KV retrieval with prefetch-based KV movement. \system effectively predicts future sparse KV demand and efficiently overlaps KV transfer with decoding computation (\cref{sec: system design}).

\item We implement \system atop of a state-of-the-art dynamic sparse KV cache system~\cite{lin2024shadowkv} and evaluate its performance on representative long-context serving workloads. The results show that \system improves decoding throughput by up to \addData{2.62$\times$} over state-of-the-art inference systems while preserving decoding latency and generation quality (\cref{sec: evaluation}).

\end{itemize}

\section{Background}
\label{sec: background}

\subsection{LLM Inference with KV Cache}

\noindent\textbf{LLM inference and batching techniques.}
Modern LLM serving systems commonly run inference on AI-specific accelerators, like GPUs, to meet the throughput and latency requirements of online applications~\cite{DBLP:conf/osdi/ZhongLCHZL0024, jain2023orca,kwon2023vllm}. A typical request is processed in two phases: prefilling, which consumes the input prompt, and autoregressive decoding, which generates output tokens one step at a time. Since decoding is sequential within each request, the end-to-end serving throughput largely depends on how many requests can be decoded concurrently~\cite{DBLP:conf/osdi/AgrawalKPMKGTR24}. Given the growing computing capability of GPUs, applications widely adopt batched decoding techniques to improve the overall token throughput and avoid wasting GPU resources. For example, a Llama-3-8B service running on \addData{a single} GPU may group \addData{4} or \addData{8} active requests into one decoding batch and process them with the same model kernels. As long as the service-level objectives (SLOs) are not violated, increasing the decoding batch size usually improves token throughput and amortizes GPU execution cost across more requests~\cite{jain2023orca, DBLP:conf/euromlsys/HeLA24}.

\noindent\textbf{KV cache and its memory footprint.}
Autoregressive decoding relies on the key-value (KV) cache to avoid repeatedly recomputing the attention states of previous tokens~\cite{zhang2023h2o, DBLP:journals/corr/abs-1911-02150}. At each decoding step, the model computes the \textit{key} and \textit{value} tensors of the newly generated token and appends them to the cache as KV entries. In later steps, the attention computation reuses these cached KV entries together with the current query to generate the next token. This reuse is essential because recomputing all historical key and value tensors would introduce substantial redundant computation, especially for long-context requests. However, the KV cache also becomes one of the dominant consumers of GPU high-bandwidth memory (HBM)~\cite{DBLP:conf/sigcomm/LiuLCRHZDY0AMHH24}. Its footprint grows with the request batch size, context length, number of layers, and model hidden dimensions. As these factors increase, especially in long-context serving, the KV cache can occupy a large fraction of the limited GPU memory capacity. Serving systems must then reduce the decoding batch size to avoid out-of-memory errors, which directly lowers token throughput and leaves expensive GPU compute resources underutilized~\cite{DBLP:conf/osdi/LeeLSS24,lin2024shadowkv, DBLP:journals/corr/abs-2403-11421}.

\subsection{Dynamic Sparse KV Cache}
\label{sec: dynamic sparse KV cache}
\begin{figure}
    \centering
    \includegraphics[width=0.9\linewidth]{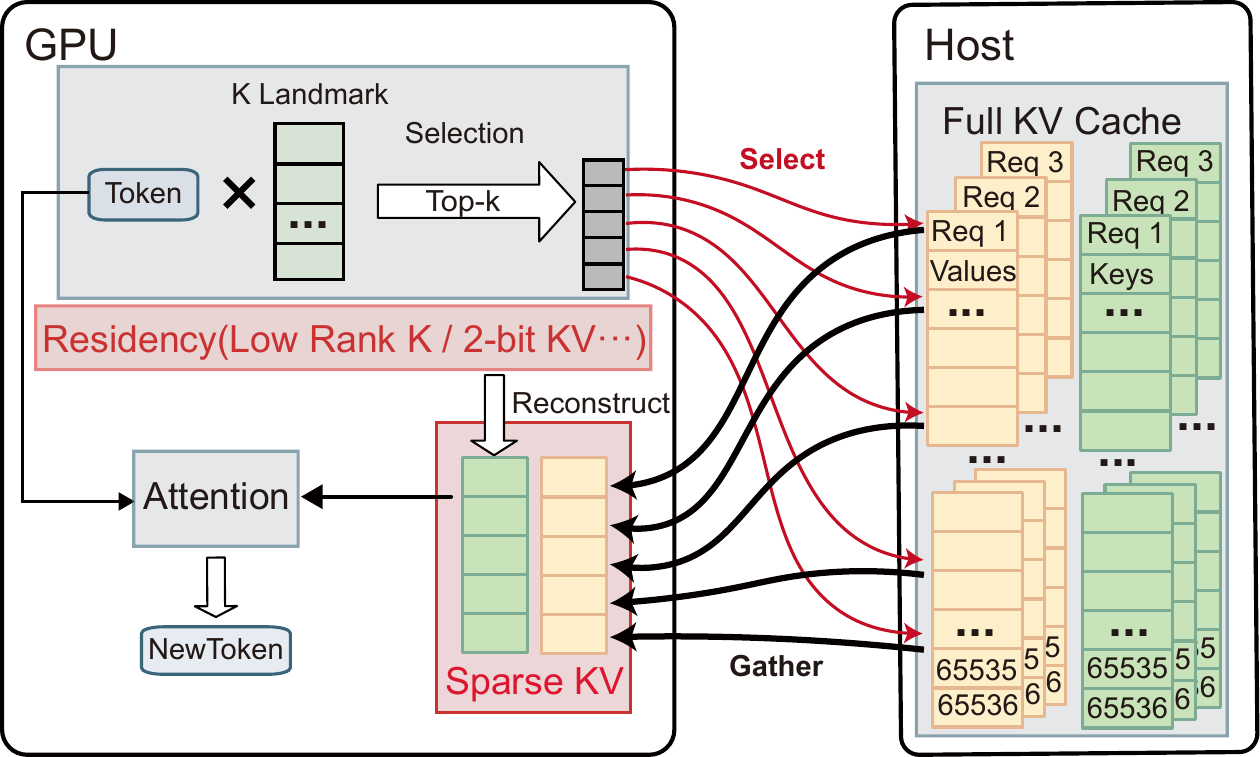}
    \caption{LLM decoding step with dynamic sparse KV cache.}
    \vspace{-0.2cm}
    \label{fig:background/dynamic_sparse}
    \vspace{-0.3cm}
\end{figure}

Dynamic sparse KV cache mechanisms reduce GPU memory usage by moving most KV entries to host memory and bringing back only the entries that are needed for each decoding step~\cite{DBLP:conf/icml/TangZZXKH24,DBLP:journals/corr/abs-2409-10516,DBLP:conf/icml/RibarCHBLO24}. The key observation behind these mechanisms is that next-token generation does not attend to all historical tokens uniformly. Instead, for many decoding steps, the attention computation depends primarily on a small subset of important historical tokens, while many cached KV entries have little impact on the generated token~\cite{xiao2023streamingllm,DBLP:conf/nips/LiHYVLYCLC24,DBLP:conf/acl/YangHGHZ024}. Based on this observation, dynamic sparse KV cache systems keep the full KV cache, or most of it, in host memory and retrieve selected KV entries into GPU HBM before attention computation, as shown in Figure~\ref{fig:background/dynamic_sparse}. The selection is typically guided by lightweight GPU-resident metadata. For example, ShadowKV~\cite{lin2024shadowkv} segments adjacent KV entries into chunks and maintains compressed landmarks for these chunks, so that the GPU can select important chunks according to the current query and then retrieve the corresponding sparse KV entries from host memory. Since the landmark representation is much smaller than the full KV cache, this approach saves GPU memory and enables a larger request batch size while still preserving access to the original KV entries when they are needed.

However, applying dynamic sparse KV cache mechanisms in practical LLM serving systems introduces nontrivial data movement overhead~\cite{DBLP:conf/osdi/LeeLSS24,DBLP:conf/eurosys/GaoCS25}. First, host-to-device bandwidth is much lower than GPU HBM bandwidth. For example, PCIe 5.0 provides only \addData{64 GBps} unidirectional bandwidth, which is far below the \addData{4.8 TBps} bandwidth level of modern GPU HBM. As a result, even fetching a sparse subset of critical KV entries can block decoding if the transfer is exposed on the critical path. Second, the selected KV entries are often scattered across the long context rather than stored as one contiguous memory region. The system therefore has to gather many noncontiguous KV entries from host memory before or during the host-to-device transfer, which further increases retrieval latency. To reduce this overhead, recent systems introduce GPU-resident auxiliary state for KV selection, reconstruction, or communication scheduling. ShadowKV~\cite{lin2024shadowkv} keeps a low-rank representation of the key cache on the GPU and reconstructs selected key entries on the fly, while SpeCache~\cite{chen2024specache} maintains a compact quantized KV representation to guide future KV prefetching. These auxiliary states help reduce exposed retrieval latency without directly evicting useful KV information. However, because they must be accessed during decoding, they also consume valuable GPU memory. In this paper, we refer to the GPU memory occupied by such auxiliary state as \textit{\residency}.

\section{Motivation}
\label{sec: motivation}
In this section, we first introduce the severe throughput \tcr{issues} when we try to apply the dynamic sparse KV cache mechanisms (\cref{sec: motivation/ throught issue}). And we then present that the throughput bottleneck stems from the memory occupancy of \residency (\cref{sec: motivation/ memory analysis}).

\subsection{Poor Serving Throughtput}
\label{sec: motivation/ throught issue}

We deploy model inference systems with novel dynamic sparse KV cache mechanisms, including ShadowKV and SpeCache, on a testbed that consists of 8 GPUs. The memory capacity of each GPU is 80 GB and the interconnections between GPU and CPU offer a bandwidth of up to \addData{64 GBps}. With this testbed, we serve popular LLMs, including the Llama series~\cite{grattafiori2024llama3herdmodels} and Qwen series~\cite{qwen2025qwen25technicalreport} models, with model sizes spanning from \addData{8B} to \addData{32B} and use mainstream long-context benchmarks such as RULER~\cite{hsieh2024rulerwhatsrealcontext} to validate serving quality.

\begin{figure}
    \centering
    \includegraphics[width=0.9\linewidth]{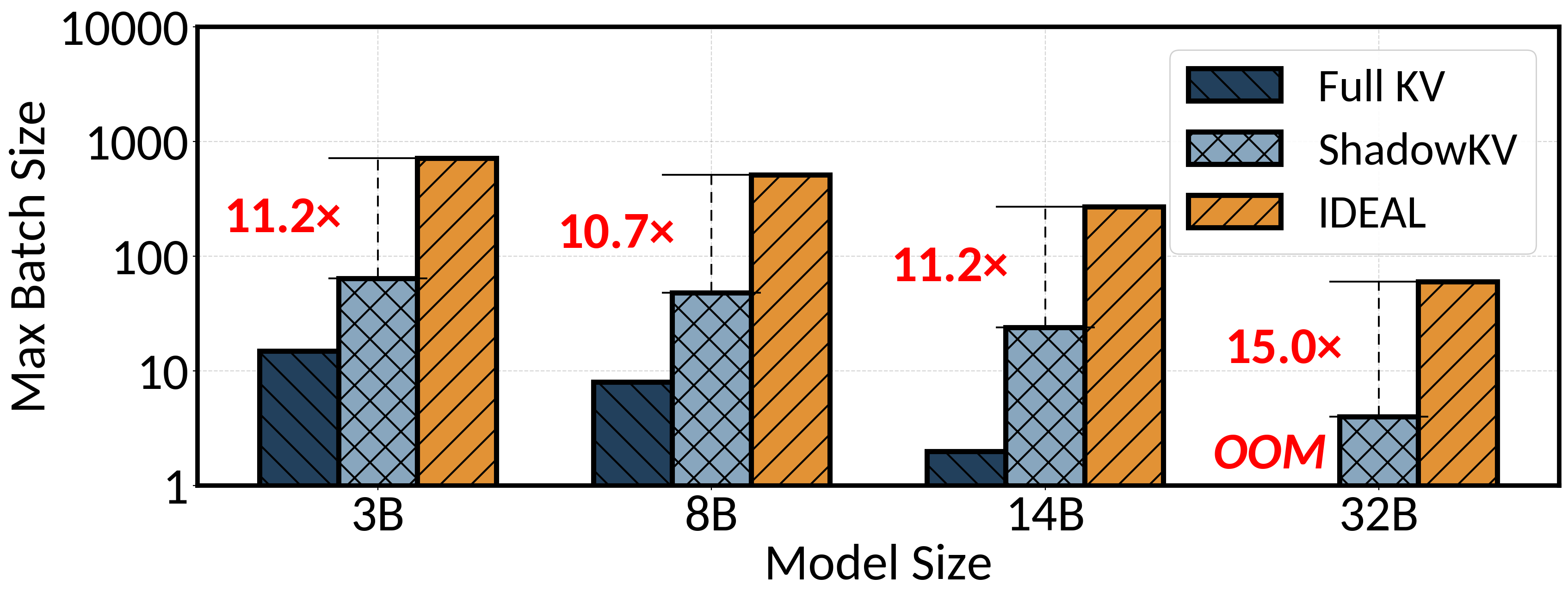}
    \caption{Maximum request batch size of decoding using full KV cache, ShadowKV and ideal scaling.}
    \vspace{-0.2cm}
    \label{fig: motivation/batch_vs_model}
    \vspace{-0.3cm}
\end{figure}

\noindent\textbf{Limited batch size.}
We first measure the maximum request batch size that each system can support without triggering out-of-memory (OOM) errors. To understand how far existing systems are from the memory capacity limit, we also compute an ideal batch size by excluding the KV cache from GPU memory while keeping model weights and necessary runtime activations. This ideal setting is not a deployable system, but it provides a useful upper bound on the serving concurrency that the GPU could support if KV cache memory were not the bottleneck. As shown in Figure~\ref{fig: motivation/batch_vs_model}, existing systems support much smaller batch sizes than this ideal bound. For example, when serving \addData{8B} models, full KV cache serving and ShadowKV support only \addData{4} and \addData{24} requests per batch, respectively, while the ideal setting can accommodate \addData{120} requests. The gap becomes even larger as the model size increases. When serving \addData{32B} models, ShadowKV can support only \addData{4} requests per batch, which is merely \addData{6.7\%} of the ideal batch size. These results show that dynamic sparse KV cache mechanisms reduce the memory footprint of KV entries, but they still leave serving concurrency far below what the hardware could otherwise support.

\begin{figure}
    \centering
    \includegraphics[width=0.9\linewidth]{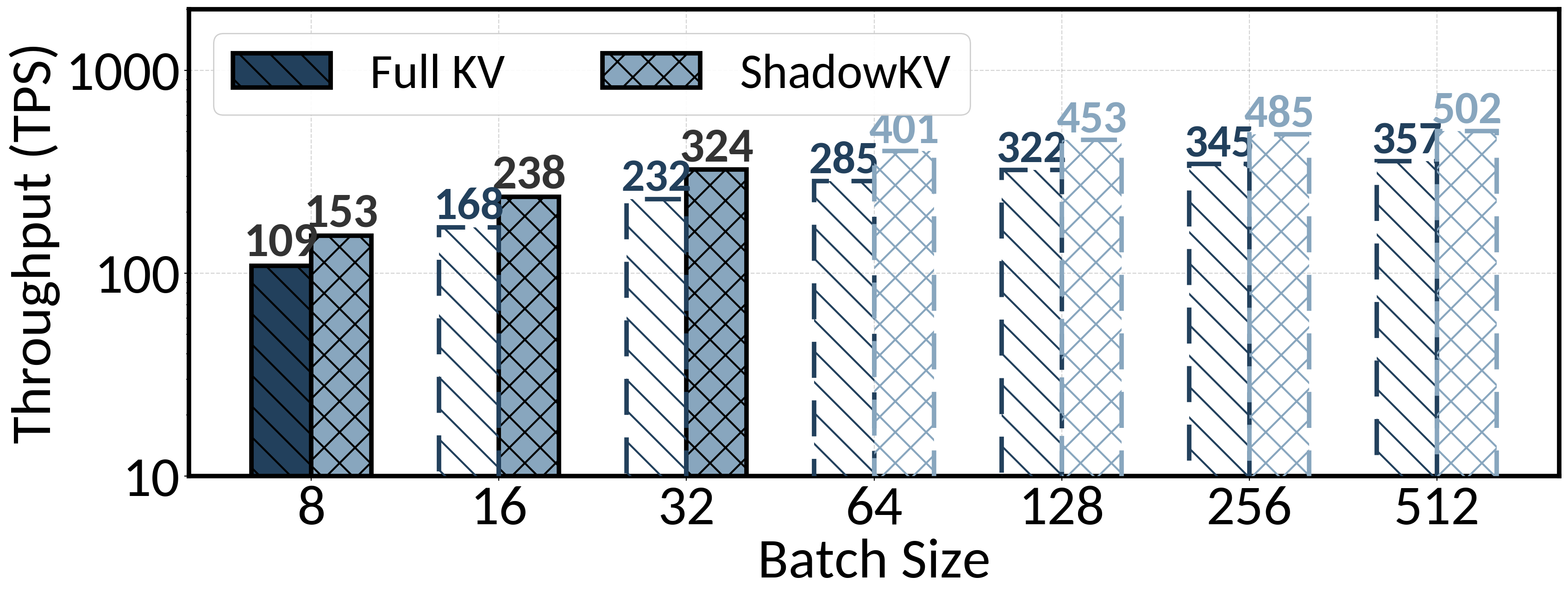}
    \vspace{-0.2cm}
    \caption{End-to-end decoding throughput.}
    \vspace{-0.4cm}
    \label{fig:motivation/throughput_vs_batch}
\end{figure}

\noindent\textbf{Throughput degradation.}
The limited batch size directly translates into lower application-level token throughput. Figure~\ref{fig:motivation/throughput_vs_batch} shows the decoding throughput of an \addData{8B} model as the request batch size increases. The throughput increases with batch size because batched decoding better amortizes GPU kernel execution and improves hardware utilization. However, existing systems stop scaling once they reach their memory capacity limit. Full KV cache serving and ShadowKV can only run up to \addData{8} and \addData{48} requests per batch, reaching \addData{52} and \addData{157} tokens per second, respectively. 
To show the throughput gaps caused by the limited batch size, we depict the ideal throughput (with the ideal batch size above) by scaling the throughput proportionally with the increase in the batch size.
Compared with this ideal throughput, full KV cache serving and ShadowKV lose \addData{72\%} and \addData{35\%} of potential token throughput, respectively. This throughput gap indicates that the batch size limitation is not merely a memory capacity issue. It directly wastes GPU compute resources and prevents long-context serving from reaching the concurrency level required by practical deployments.

\subsection{Memory Capacity Analysis}
\label{sec: motivation/ memory analysis}
\begin{figure}
    \centering
    \includegraphics[width=0.9\linewidth]{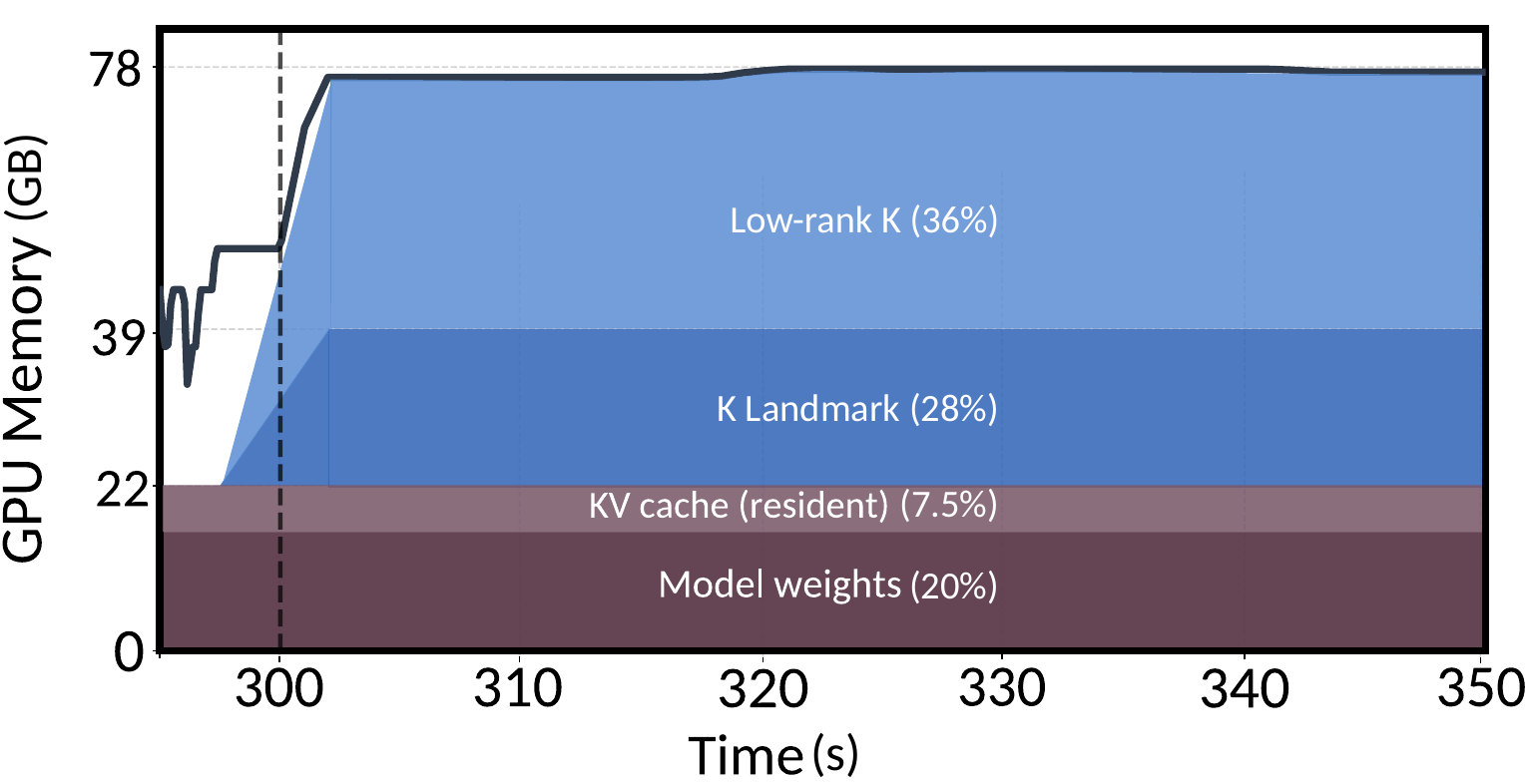}
    \vspace{-0.2cm}
    \caption{Memory footprint over time for a typical inference process.}
    \label{fig:memory_breakdown2}
    \vspace{-0.3cm}
\end{figure}

\begin{figure}
    \centering
    \includegraphics[width=0.9\linewidth]{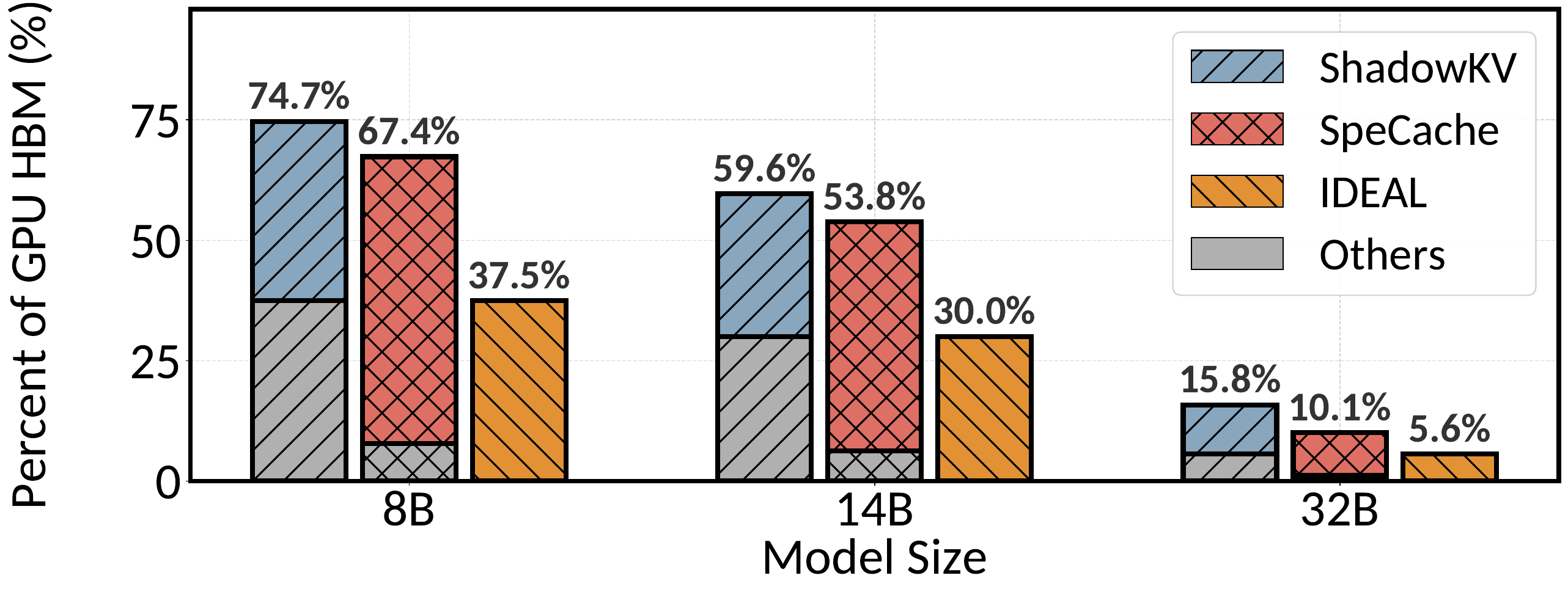}
    \vspace{-0.3cm}
    \caption{Memory occupation of residency over model sizes.
    }
    \vspace{-0.3cm}
    \label{fig:motivation/residency percent}
\end{figure}

We hereby investigate the GPU memory footprint of the evaluated inference systems to identify the root cause of the serving throughput bottleneck. Figure~\ref{fig:memory_breakdown2} breaks down the runtime GPU memory usage of a representative \addData{8B} model serving workload into \addData{model weights}, sparse KV entries, K landmarks, and \residency. The key observation is that K landmarks and \residency occupy up to \addData{$64\%$} of GPU memory and can be \addData{$8.5\times$} larger than the sparse KV entries that are actually kept in GPU HBM. This large footprint is not transient. It remains resident throughout decoding because the auxiliary state must be accessed at every decoding step for KV selection, reconstruction, or prefetching. Moreover, its size grows with request batch size and context length, which makes it increasingly difficult for dynamic sparse KV cache systems to scale to large serving workloads. To further validate this trend, we measure the fraction of GPU memory consumed by \residency across different models and systems. As shown in Figure~\ref{fig:motivation/residency percent}, \residency consistently accounts for at least \addData{$30\%$} of GPU memory in all evaluated settings. These results show that although dynamic sparse KV cache systems reduce the memory footprint of GPU-resident KV entries, they introduce a new form of memory pressure through GPU-resident auxiliary state, which becomes the limiting factor for request batch size and serving throughput.

As we presented in \cref{sec: dynamic sparse KV cache}, this large \residency is a direct consequence of reducing CPU to GPU KV retrieval overhead. Since fetching sparse KV entries from host memory over PCIe is expensive, recent systems avoid relying solely on raw data transfer. Instead, they keep auxiliary state in GPU HBM to accelerate KV selection, reduce transferred data volume, or reconstruct part of the sparse KV cache on the fly. For example, ShadowKV keeps a low-rank key cache on the GPU to reconstruct selected key entries during decoding and fetches only the corresponding value entries from host memory. SpeCache similarly keeps a compact quantized KV representation to guide future KV prefetching. These designs effectively reduce exposed retrieval latency, but they shift part of the KV cache cost back to GPU memory. The resulting \residency is nontrivial becau se it scales with model depth, batch size, context length, hidden dimension, sparse budget, chunk size, and quantization or low-rank configuration. As models and contexts grow, this auxiliary state can consume a substantial portion of the HBM capacity that dynamic sparse KV cache mechanisms are intended to save. Therefore, the throughput bottleneck in existing systems is no longer only the sparse KV entries themselves, but the GPU-resident auxiliary state required to make sparse retrieval fast enough for decoding.

\section{Predictive KV Retrieval}
\label{sec: Predictive KV Retrieval}

In this section, our goal is to eliminate \residency while preserving the efficiency of dynamic sparse KV cache. To this end, we first show that sparse KV retrieval exhibits strong predictability across consecutive decoding steps (\cref{sec: MTP/predictability}). We then explain how this predictability creates an opportunity to prefetch selected KV entries from host memory before they are consumed (\cref{sec: opportunities}), and discuss the practical challenges of turning this opportunity into an efficient decoding pipeline (\cref{sec: challenges}).
\subsection{KV Retrieval Predictability}
\label{sec: MTP/predictability}

In dynamic sparse KV cache systems, each decoding step computes a retrieval index that specifies which KV entries should be fetched from host memory for attention computation. This retrieval index is determined by the current query and the GPU-resident landmarks that summarize the offloaded KV entries. More concretely, in the decoding step $S_{i}$, which is responsible for decoding token $T_{i+1}$, the KV index of token $T_{i+1}$ can be computed by token $T_{i}$ and the landmarks.
This dependency creates a natural opportunity for prefetching. If token $T_i$ were known during step $S_{i-1}$, the system could compute the retrieval index for $T_{i+1}$ ahead of time and fetch the corresponding KV entries before step $S_i$ starts. The landmarks do not prevent this precomputation because they are generated before decoding and remain fixed throughout the decoding process. The real obstacle is the token dependency of autoregressive decoding. Token $T_i$ is not available until step $S_{i-1}$ completes, which makes exact retrieval index precomputation infeasible.

\begin{figure}
    \centering
    \includegraphics[width=0.9\linewidth]{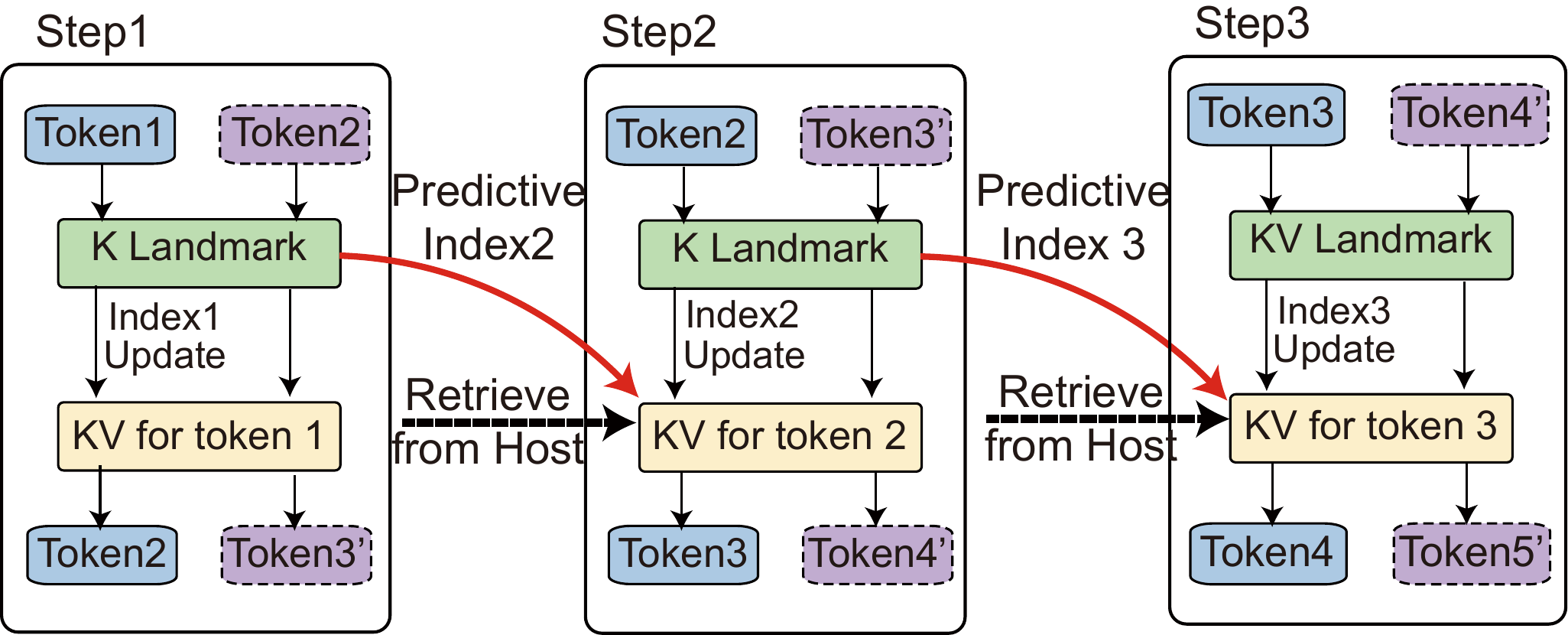}
    \vspace{-0.3cm}
    \caption{KV retrieval prediction via \spectoken.}
    \label{fig:dual_decode}
    \vspace{-0.4cm}
\end{figure}

\begin{figure}[!tb]
    \centering
    \subfigure[KV and token prediction.]{
        \label{fig: MTP/pred_acc_vs_token}
        \hspace{-0.28cm}
        \includegraphics[width=0.48\linewidth]{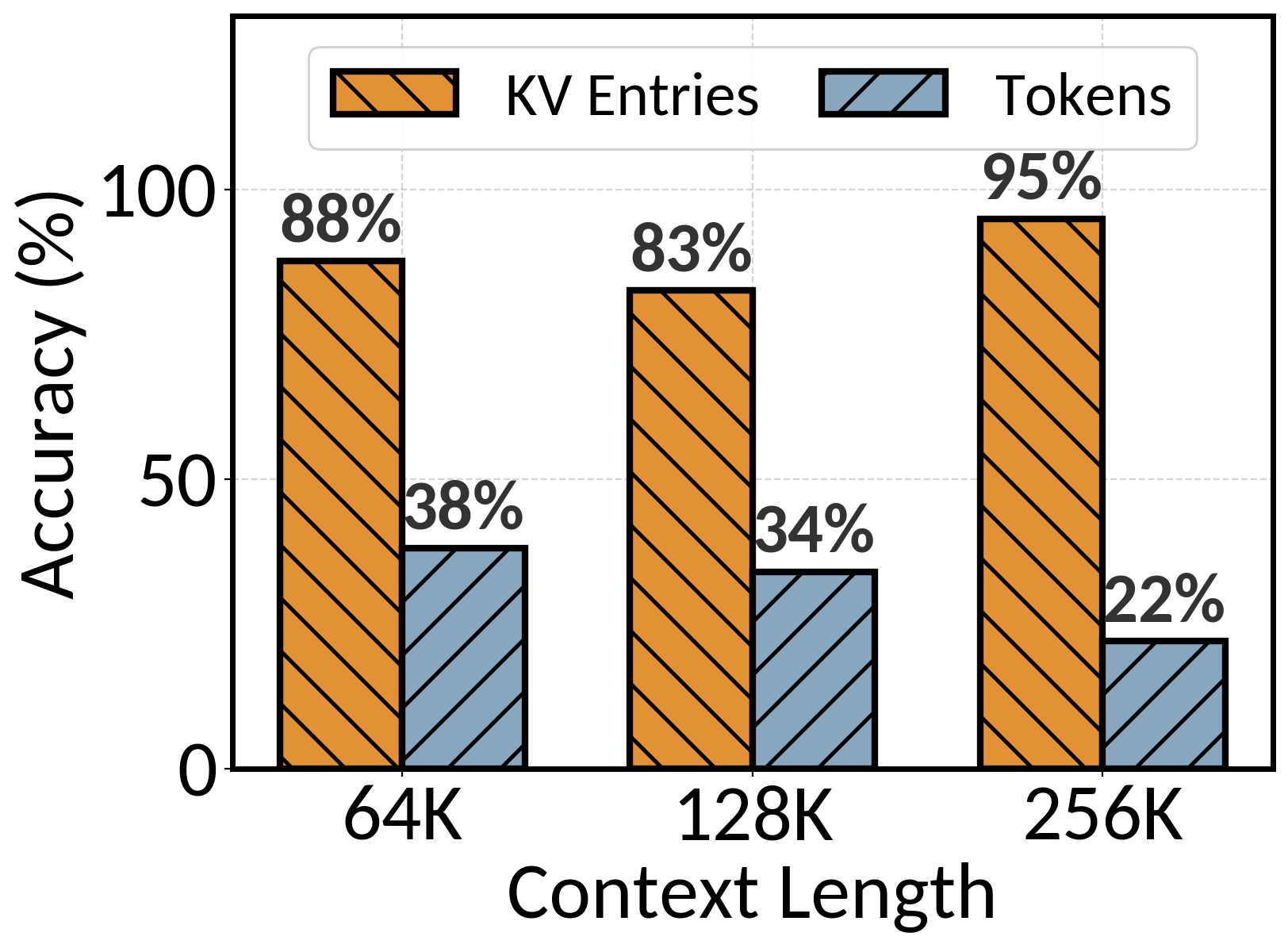}
    }
   \hspace{-0.2cm}
    \subfigure[Initial \spectoken selection.]{
        \label{fig: pred_acc_vs_random}
        \includegraphics[width=0.48\linewidth]{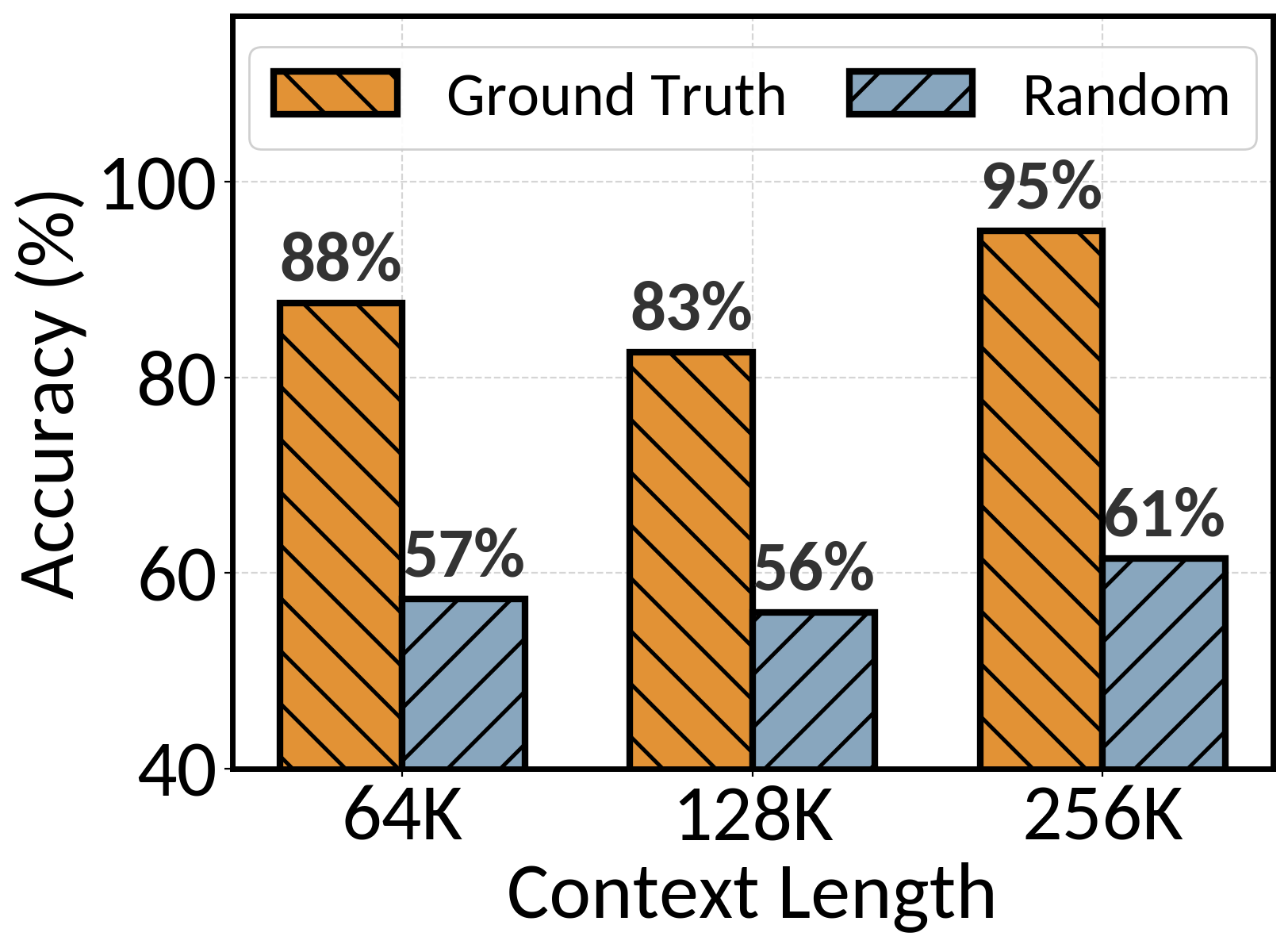}
    }
    \vspace{-0.3cm}
    \caption{Accuracy of KV prediction.}
    \vspace{-0.4cm}
\end{figure}

\noindent\textbf{Retrieval index prediction.}
To overcome this dependency, we use a lightweight token prediction mechanism to estimate the retrieval index of the next decoding step. 
Recent studies~\cite{DBLP:conf/asplos/XuPWZY0026} provide approaches like speculative decoding to \tcr{precisely predict the decoding tokens}. However, they either introduce heavy computation and memory overhead by running a draft model, which precomputes the tokens~\cite{DBLP:conf/naacl/YanAV25,schuster2023lookahead,leviathan2023speculative}, or only cover partial model layers~\cite{DBLP:conf/acl/ElhoushiSLHWL0A24,DBLP:conf/isca/XuPZCLLWD25,DBLP:conf/icml/GloeckleIRLS24}, which makes them not fit in our case.
Our insight is that we do not need to predict the exact next token. \textit{A rough token prediction is enough for predicting the correct KV cache index}. To show this, we introduce \textit{\spectoken} into decoding, as shown in Figure~\ref{fig:dual_decode}. In each decoding step, the model generates both the output token and a \spectoken that is placed immediately after the output token in the sequence. Because the \spectoken is computed with the output token included in its causal context, it carries similar positional and contextual information as the next true token. Thus, in each decoding step, we can predict the KV cache index of the next token, by computing it with the landmarks and the \spectoken. Figure \ref{fig: MTP/pred_acc_vs_token} presents the prediction accuracy of both \spectoken and next KV cache. We can observe that for token prediction, the accuracy can be low during decoding steps. However, compared to that, the KV prediction accuracy can always be higher, with an average accuracy of \addData{88\%} across the serving configurations. 

\noindent\textbf{Sensitivity to the first \spectoken.}
We further study how the initial \spectoken affects retrieval index prediction. After the first \spectoken is chosen, subsequent \spectokens are generated by the model itself, so the initial token can strongly influence the prediction trajectory. We evaluate two initialization strategies for the first \spectoken $T'_2$, including using the ground truth output token $T_2$ and using a randomly selected token. As shown in Figure~\ref{fig: pred_acc_vs_random}, using $T_2$ as the first \spectoken consistently yields much higher retrieval index prediction accuracy. In contrast, when adopting random initialization, the accuracy can degrade significantly to even \addData{56\%} on average. This result shows that retrieval index prediction is sensitive to the first \spectoken, and using ground-truth tokens can obtain the best performance.

\begin{figure*}
    \centering
    \includegraphics[width=0.9\linewidth]{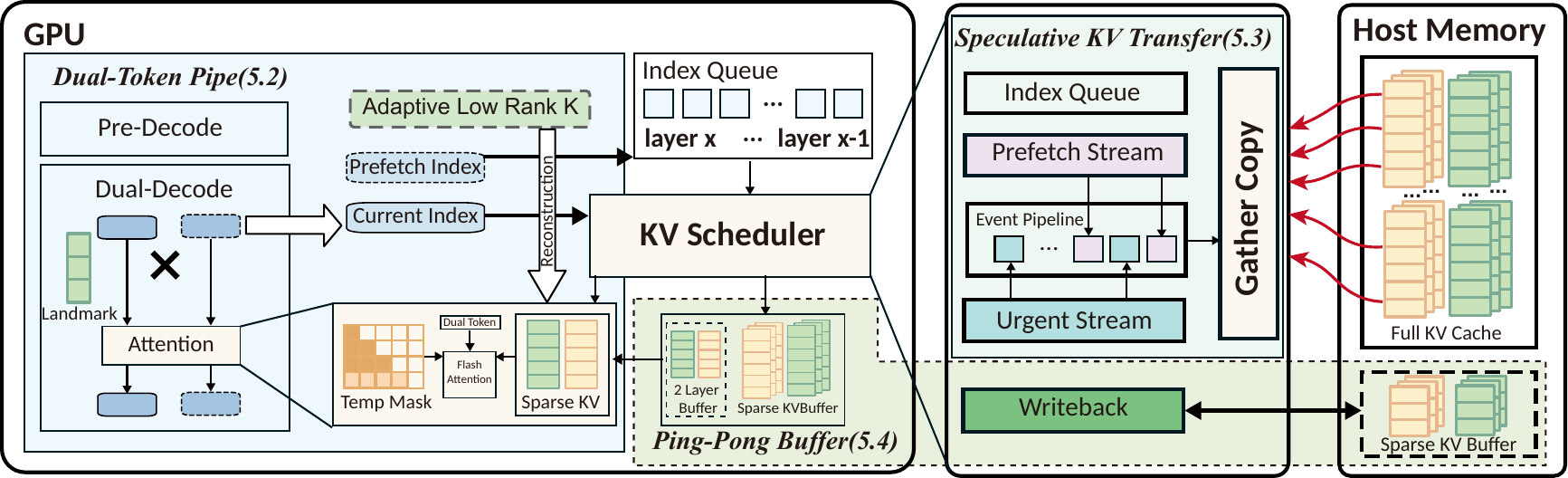}
    \vspace{-0.3cm}
    \caption{Overview of \system}
    \vspace{-0.4cm}
    \label{fig:design/overview}
\end{figure*}

\subsection{Opportunities from Predictability}
\label{sec: opportunities}

The predictability of KV retrieval indices creates a direct opportunity to move sparse KV retrieval out of the critical path of decoding. 
Thus, we can directly apply it during decoding and prefetch the predicted KV cache of the next step, which we call the \textit{speculative KV cache}.
This prefetching offers a direct opportunity to move sparse KV retrieval out of the critical path of decoding. Since they are not consumed by the current decoding step, their host-to-device transfer can be issued early and overlapped with useful GPU computation. This allows us to hide much of the CPU-to-GPU KV transfer latency behind the computation of the current step, instead of waiting until the true retrieval index is available and then fetching the selected KV entries on the critical path. In this way, predictive retrieval can reduce exposed communication latency without changing the attention result, as any missing entries can still be corrected before attention computation.

This opportunity also addresses the memory bottleneck identified in \cref{sec: motivation/ memory analysis}. Existing dynamic sparse KV cache systems rely on \residency because they need GPU-resident auxiliary state to accelerate KV selection, reconstruction, or future prefetching. Predictive retrieval provides a different way to reduce retrieval latency. Rather than keeping a low-rank key cache, a quantized KV cache, or other large auxiliary state in GPU HBM, we can use the predicted retrieval index to start transferring the likely needed KV entries earlier. This shifts the optimization target from storing more auxiliary state on the GPU to scheduling KV movement earlier in the decoding pipeline. 
As a result, we can potentially preserve the latency benefit of sparse KV retrieval while releasing a large fraction of GPU memory previously occupied by \residency. 

\subsection{Challenges}
\label{sec: challenges}
Although predictive KV retrieval provides a promising way to reduce exposed CPU-to-GPU transfer latency and remove large \residency, applying this idea in a real decoding pipeline faces critical challenges.
\noindent\textbf{C1: Balancing prefetch latency and buffer footprint.}
Speculative prefetching is only beneficial when KV transfer is scheduled at the right time. If prefetching starts too late, the selected KV entries may not arrive before attention consumes them, and decoding still stalls on host-to-device transfer. If prefetching starts too early, these entries must remain in GPU HBM for longer, increasing the sparse KV buffer footprint and reducing the batch size benefit of removing \residency. 

\noindent\textbf{C2: Controlling the computational overhead of \spectoken generation.}
The \spectoken should help predict future KV entries, but it should not make each decoding step much more expensive. A naive implementation may run another decoding pass or launch separate kernels for the output token and the \spectoken, which increases computation, kernel launch overhead, and HBM traffic. Such overhead can easily cancel the latency benefit of KV prefetching. 

\noindent\textbf{C3: Bounding the prefetched KV buffer.}
Predictive retrieval reduces the need for \residency, but it can still consume large GPU memory if prefetched KV entries are buffered naively. A straightforward implementation would reserve GPU HBM for the speculative KV cache of future decoding work. This buffer grows with batch size, context length, and sparse KV budget, and can quickly consume the memory saved by removing auxiliary state. 

\section{\system Design}
\label{sec: system design}
In this section, we first introduce the overall design of \system (\cref{sec: system overview}). We then describe the three key components in detail: dual-token decoding pipeline (\cref{sec: Dual-Token Decoding Pipeline}), speculative KV cache transfer scheduler (\cref{sec: Speculative KV Cache Transfer Schedule}), and layer-scoped KV memory management (\cref{sec: Layer-Scoped KV Memory Management}).

\subsection{System Overview}
\label{sec: system overview}

Our goal is to improve long-context serving throughput without increasing decoding latency or reintroducing large GPU-resident auxiliary state. To this end, we design \system, a predictive KV retrieval system that uses retrieval index prediction to prefetch sparse KV entries from host memory before they are needed by attention. As shown in Figure~\ref{fig:design/overview}, \system consists of three components that jointly address the challenges in \cref{sec: challenges}.

\noindent\textbf{Dual-token decoding pipeline.}
\system first introduces a dual-token decoding pipeline to generate the output token and the \spectoken within the same decoding step. Instead of running an additional decoding pass for token prediction, \system co-executes the two tokens in the single model kernels and uses a causal mask to preserve the normal autoregressive dependency. This allows the \spectoken to provide retrieval guidance with little additional execution overhead. To stabilize retrieval index prediction from the beginning of decoding, \system initializes the first \spectoken with the real output token produced by a lightweight pre-decoding step and reuses the KV entries generated in this step.

\noindent\textbf{Speculative KV cache transfer schedule.}
Given the retrieval index predicted from the \spectoken and landmarks, \system prefetches the corresponding sparse KV entries from host memory to GPU HBM before the next attention computation consumes them. The transfer schedule is designed to hide CPU-to-GPU latency while avoiding excessive early buffering. Meanwhile, \system handles prediction misses by giving missing KV entries higher transfer priority than background prefetches and by selectively reconstructing key entries when fetching them would otherwise block decoding. 

\noindent\textbf{Layer-scoped KV memory management.}
\system manages the GPU-resident sparse KV working set with a layer-scoped memory manager. It uses a two-layer ping-pong buffer so that one buffer can serve the current attention computation while the other receives prefetched KV entries. 
To reduce host-side gather overhead, \system further uses a GPU-side sparse KV assembly kernel that places transferred KV entries into the attention layout in GPU HBM. This design keeps the prefetched KV buffer small while maintaining correct and efficient sparse attention execution.

\subsection{Dual-Token Decoding Pipeline}
\label{sec: Dual-Token Decoding Pipeline}
As discussed in \cref{sec: MTP/predictability}, retrieval index prediction requires a \spectoken that approximates the next token's attention behavior. A straightforward implementation is to run an extra decoding pass to generate this token in addition to the normal output token. However, this approach is too expensive for serving. It nearly doubles the per-step model execution, introduces additional kernel launches, and increases GPU HBM traffic. Moreover, \spectokens are not independent predictions. Each \spectoken is generated from the previous output token and the previous \spectoken, so a poor initialization can propagate through later decoding steps and reduce retrieval index prediction accuracy. To address these issues, \system designs a dual-token decoding pipeline that co-executes the output token and the \spectoken within the normal decoding flow and initializes the first \spectoken with the real output token.

\noindent\textbf{Dual-token decoding.}
Modern GPUs are most efficient when model kernels operate on sufficiently large tensor shapes. \system leverages this property by computing the output token and the \spectoken together instead of launching two separate decoding passes. At decoding step $S_i$, \system takes the current input token $T_i$ and the speculated token $T'_{i+1}$ as the two token inputs. A naive batching implementation can concatenate them along the batch dimension, allowing kernels such as GeMM to process them concurrently. 
However, as $T_{i}$ and $T_{i+1}^{'}$ show causality, batching them can make the \textit{attention} computation, which models the sequential relationship, ignore their correlations and affect the token prediction. Therefore, \system proactively transforms the input tensor shape to reconcatenate both types of tokens that belong to the same requests, along the sequence dimension. Meanwhile, following the basic decoding logic, the previous tokens can never be affected by the following tokens, e.g., $T_{i}$ will not be affected by $T_{i+1}^{'}$. To achieve this, we add a mask that prevents the output token from using the data of the speculative token in the attention computation.

\noindent\textbf{First \spectoken initialization.}
The first \spectoken is important because it determines the initial trajectory of subsequent speculative tokens. As shown in \cref{sec: MTP/predictability}, using the real second token $T_2$ as the initial \spectoken produces much more accurate retrieval index prediction than using a random token. Therefore, before entering the dual-token pipeline, \system performs one lightweight pre-decoding step with the first decoding input token $T_1$ to generate $T_2$\footnote{$T_1$ is the first decoding input token produced after prefilling.}. \system then uses $T_2$ as $T'_2$ and starts dual-token decoding from the following execution. This initialization adds only one extra decoding step over an entire generation process that usually contains many decoding steps. In addition, the pre-decoding step has already fetched or prepared the KV entries needed for its execution, and \system reuses these entries when entering the dual-token pipeline, further reducing the effective startup overhead.

\begin{figure}
\centering
  \subfigure[Prioritized KV transfer schedule.]{
  \label{fig:design/normal_timeline}
  \begin{minipage}{0.4\textwidth}
  \centering
    \includegraphics[scale=0.4]{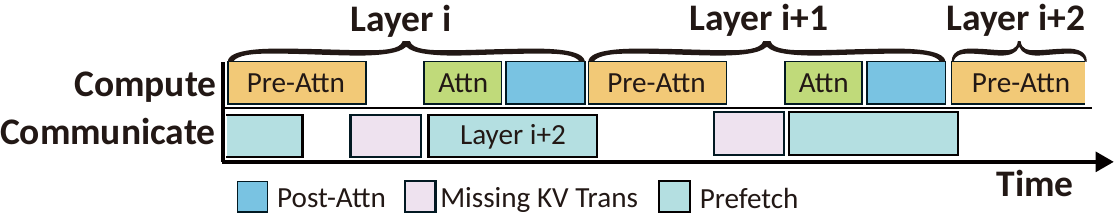}
  \end{minipage}
  }\\
  \vspace{-0.2cm}
  \subfigure[KV transfer overlapping with adaptive K reconstruction]{
  \label{fig:design/adaptive_residency}
  \begin{minipage}{0.4\textwidth}
  \centering
    \includegraphics[scale=0.4]{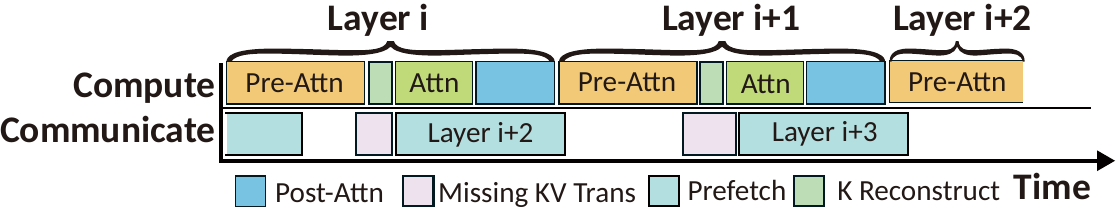}
  \end{minipage}
  }
    \vspace{-0.4cm}
   \caption{\system layer-aware KV transfer overlapping.}
  \label{fig:design/timeline} 
  \vspace{-0.4cm}
\end{figure}

\subsection{Speculative KV Cache Transfer Schedule}
\label{sec: Speculative KV Cache Transfer Schedule}
With the dual-token decoding pipeline, each decoding step can use the \spectoken and landmarks to predict the retrieval index of the next step. However, prediction alone does not remove KV retrieval latency. The selected KV entries still have to be gathered from host memory and transferred to GPU HBM before attention consumes them. Moreover, prediction is not always exact, so the entries missed by the predicted retrieval index must also be fetched or reconstructed to preserve the same sparse attention result. To address these issues, \system designs a speculative KV transfer schedule that overlaps predicted KV retrieval with decoding computation, prioritizes missing KV entries on the critical path, and selectively reconstructs key entries for layers with low prediction accuracy.

\noindent\textbf{Layer-aware KV transfer overlapping.}
The speculative KV cache contains the KV entries selected by the predicted retrieval index. A naive implementation would fetch all speculative KV entries for the next decoding step as early as possible. This can hide communication latency, but it also keeps prefetched entries in GPU HBM for a long time and increases the sparse KV buffer footprint. \system instead schedules KV transfer according to the layer execution order of the model. Although the retrieval index is predicted in the previous decoding step, \system does not immediately fetch all selected entries. For a target layer $i+2$, \system starts prefetching its speculative KV entries after the attention computation of layer $i$ completes, as shown in Figure~\ref{fig:design/normal_timeline}. This timing provides a useful overlap window for CPU-to-GPU transfer while avoiding long-lived prefetched buffers. If the transfer is delayed until layer $i+1$ finishes attention, layers with larger sparse KV budgets may not receive their entries in time and the following attention computation has to wait. In this way, layer-aware scheduling balances transfer latency hiding and GPU memory usage. We describe the corresponding buffer management in \cref{sec: Layer-Scoped KV Memory Management}.

\begin{figure}
    \centering
    \includegraphics[width=0.9\linewidth]{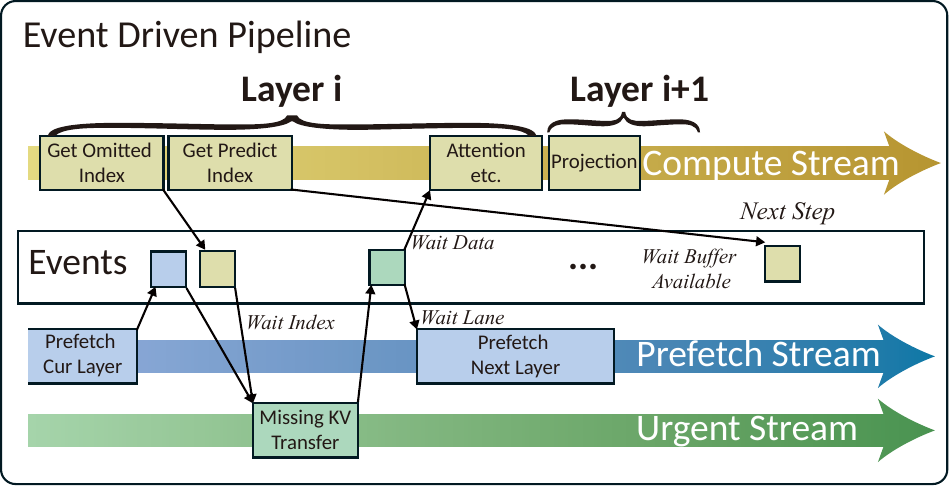}
    \vspace{-0.3cm}
    \caption{Streams that ensure prioritized transmission.}
    \label{fig:Events}
    \vspace{-0.4cm}
\end{figure}

\noindent\textbf{Prioritized missing KV transfer.}
Although the predicted retrieval index covers most required KV entries, prediction misses are inevitable. These missing KV entries must be available before attention starts, otherwise the sparse attention result would differ from the result computed with the true retrieval index. The difficulty is that the true retrieval index is only known close to the attention computation, leaving little time for correction. If missing entries are placed in the same transfer queue as background prefetches, they may wait behind future speculative transfers and directly stall decoding. \system therefore separates urgent correction from background prefetching. As shown in Figure \ref{fig:Events}, missing KV entry transfers are issued to a high-priority urgent stream, while speculative prefetches are issued to a lower-priority prefetch stream. This priority separation allows corrections for the current attention computation to finish first, while still keeping background prefetches progressing when bandwidth is available.

\noindent\textbf{Adaptive key reconstruction.}
For some layers, the predicted retrieval index has lower accuracy and produces more missing KV entries. Fetching all missing keys and values from host memory in these layers can still expose noticeable transfer latency. \system therefore adaptively enables key reconstruction for such layers, as shown in Figure \ref{fig:design/adaptive_residency}. The key idea follows the observation that selected key entries can be reconstructed from a low-rank key cache kept on the GPU~\cite{lin2024shadowkv}, while value entries are still fetched from host memory. \system applies this reconstruction only to layers where profiling shows consistently low retrieval prediction accuracy. This selective use avoids fetching many missing key entries on the critical path, while keeping the additional \residency small because only a small number of layers require the low-rank key cache. The layer set can be determined offline through lightweight profiling on representative long-context workloads.

\subsection{Layer-Scoped KV Memory Management}
\label{sec: Layer-Scoped KV Memory Management}
Efficient memory management is critical to making speculative KV prefetching practical under long context serving. Although the previous section hides most CPU-to-GPU transfer latency by overlapping speculative KV movement with layer execution, a naive implementation would still allocate sparse KV buffers for many transformer layers at the same time. This design quickly becomes memory-inefficient because the sparse KV working set grows with the batch size, context length, number of selected KV entries, and model depth. The challenge is that \system must provide enough GPU-resident KV space for asynchronous communication while ensuring that the attention kernel always observes a complete and correct sparse KV set for the current layer. Allocating too few buffers can introduce data hazards between computation and communication, while allocating per-layer buffers across the whole model would largely offset the memory benefit of KV offloading. To address this problem, \system introduces a layer-scoped KV memory manager that maintains only two active sparse KV buffers on the GPU and coordinates their reuse across transformer layers with CUDA stream events.

\begin{figure}
    \centering
    \includegraphics[width=0.9\linewidth]{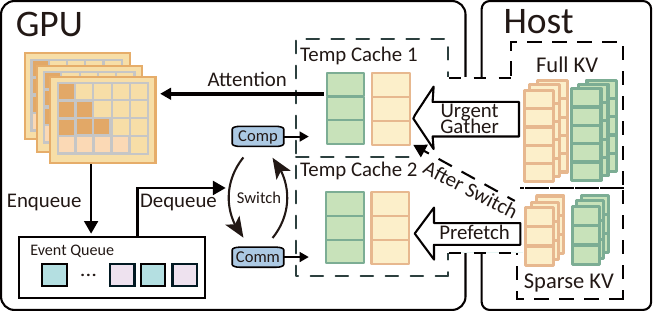}
    \vspace{-0.3cm}
    \caption{\system two-layer ping-pong buffer.}
    \label{fig:design/pingpong}
    \vspace{-0.4cm}
\end{figure}

\noindent \textbf{Two-layer ping-pong buffering.}
\system uses a two-layer ping pong buffer to decouple the sparse KV working set of the current layer from the speculative KV transfer of a future layer. At any time, one buffer serves as the compute buffer that is consumed by the attention kernel of layer $i$, while the other buffer serves as the transfer buffer that receives prefetched KV entries for layer $i+2$. As we demonstrate in Figure\ref{fig:design/pingpong}, After the attention computation of layer $i$ finishes, \system switches the roles of the two buffers so that the released compute buffer can be reused as the transfer destination for a later layer. This reduces the GPU resident KV buffer from a full model depth allocation to a constant two-layer allocation. To ensure correctness, \system explicitly inserts CUDA events between the compute stream and the communication streams. As shown in Figure\ref{fig:Events}Before launching attention for a layer, the compute stream waits for the event that marks the completion of all required KV transfers for that layer, including both speculative prefetches and urgent missing KV updates. Similarly, before the communication stream writes into a buffer, it waits for the event that confirms the previous attention computation has released that buffer. These event dependencies prevent read-after-write and write-after-read hazards without forcing global synchronization across all streams. As a result, the ping pong buffer preserves the same sparse attention semantics as a larger per-layer buffer design, but with substantially lower GPU memory residency.

\noindent \textbf{GPU-side sparse KV assembly.} Sparse KV retrieval naturally requires gathering many noncontiguous KV entries according to the selected indices, but performing this gather on the CPU is inefficient because it creates a large number of small and irregular memory copies before the data can be sent over PCIe. This CPU-side gather becomes especially expensive at large batch sizes, where the sparse KV working set contains many scattered entries from different requests, heads, and chunks. \system therefore adopts a GPU-controlled GPU memory update mechanism. Instead of asking the CPU to assemble the final sparse KV layout, the host-side transfer path only moves selected KV blocks into a GPU staging region, and \system launches lightweight GPU kernels to compact, reorder, and merge them into the active sparse KV buffer. The GPU update kernel uses the current layer index, the predicted prefetch index, and the urgent missing index to place each KV entry at its final attention layout inside HBM. This design shifts the irregular gather and layout update from CPU memory to GPU memory, where massive parallelism and higher memory bandwidth make the operation much cheaper. It also enables deferred reordering. Speculative entries can be written into temporary slots first, and only after the ground truth index is resolved does \system update the final sparse KV layout for attention. 

\section{Evaluation}
\label{sec: evaluation}

In this section, we first describe the overall evaluation setup (\cref{subsec:setup}). We then demonstrate that \system outperforms state-of-the-art systems by up to \adddata{$2.62\times$} in decoding throughput (\cref{subsec:e2eperformance}). Finally, we show more results on prediction accuracy (\cref{sec: Accuracy Analysis}) and the modular studies (\cref{sec: Modular Study}).

\subsection{Experiment Setup}
\label{subsec:setup}

\noindent\textbf{Testbed.}
We conduct our experiments on a server which equipped with 8 Nvidia Hopper GPUs, where a single GPU provides 80GB of memory capacity (HBM). The host side has two INTEL(R) XEON(R) PLATINUM 8558 processors (96 cores in total), with an aggregated host memory of 1.6TB. The GPUs and CPUs are interconnected via a PCIe 5.0 bus that provides a bidirectional bandwidth of up to 128GB/s. Pytorch-2.3.1 and CUDA-12.8 are adopted in our platform. We implement \system atop ShadowKV, which is one of the state-of-the-art systems for sparse KV cache inference.

\noindent\textbf{Baseline and workloads.} We compare \system against three state-of-the-art LLM serving systems, including vLLM, SpeCache and ShadowKV. vLLM adopts a conventional full KV-cache method in which the entire KV matrices reside in GPU memory. In contrast, SpeCache and ShadowKV employ dynamic sparse KV cache techniques to reduce GPU memory usage. Our evaluation covers models from the Llama-3 and Qwen-2.5 families, with parameter sizes ranging from 8B to 32B. To evaluate long-context serving performance, we use RULER~\cite{}, a widely adopted benchmark for long-context LLMs. The input context length varies from \addData{64K} to \addData{512K} tokens. Because RULER does not provide request timestamps, we follow the common practice in previous works~\cite{hsieh2024rulerwhatsrealcontext} and synthesize request arrivals using a Poisson process. We vary the request arrival rate from \addData{0.1} to \addData{100} requests per second, allowing us to evaluate system behavior under different workload intensities.

\begin{figure}
    \centering
    \includegraphics[width=0.9\linewidth]{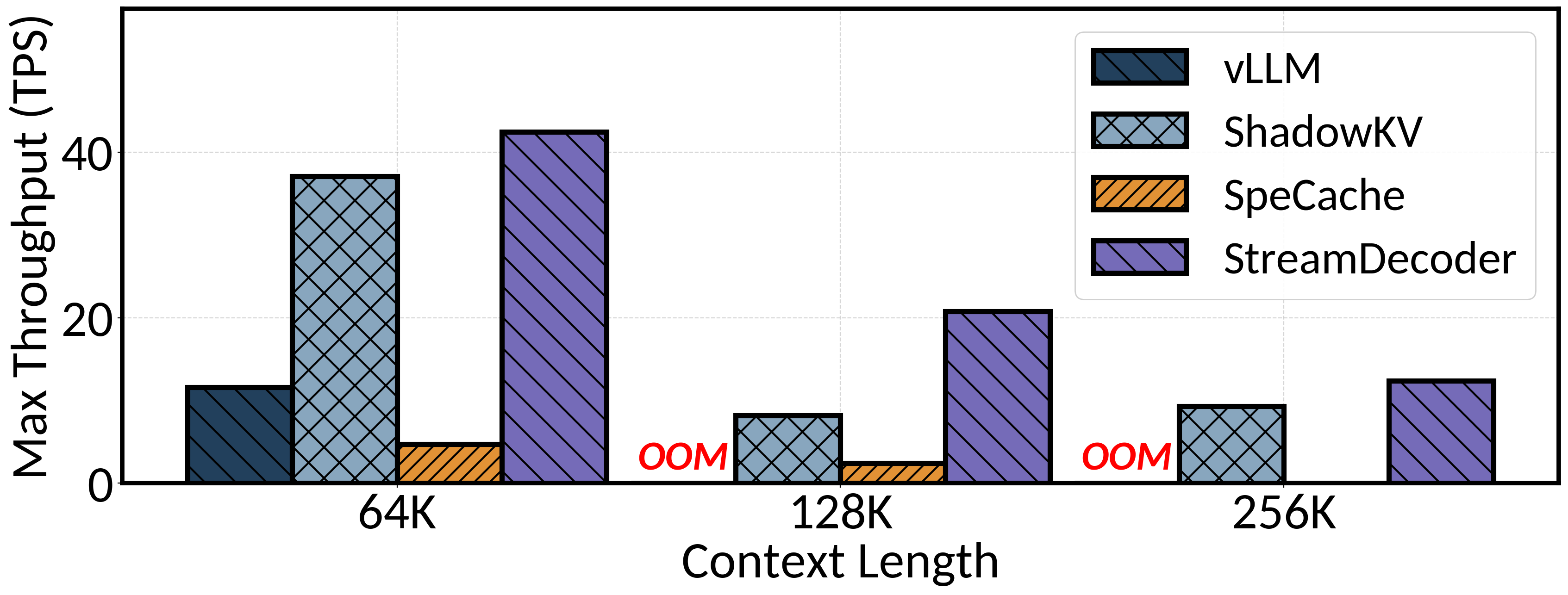}
    \vspace{-0.3cm}
    \caption{End-to-end serving throughput.}
    \vspace{-0.4cm}
    \label{fig:eval/throughput}
\end{figure}

\noindent\textbf{Metrics.} The objective of \system is to maximize the effective decoding batch size for long-context LLM serving. Accordingly, our evaluation focuses on two key performance metrics: serving throughput and request completion time (RCT), while maintaining latency comparable to existing systems. We further evaluate the KV cache prediction accuracy to validate our key insight on KV predictability. Meanwhile, we measure GPU memory usage and decoding latency to quantify the effectiveness of \system in reducing memory footprint and to assess the contribution of its individual design components.

\subsection{End-to-End Performance}
\label{subsec:e2eperformance}
\textbf{Decoding throughput improvement.} We evaluate the end-to-end decoding throughput of \system across context lengths from 64K to 512K using Qwen-2.5-32B. Figure~\ref{fig:eval/throughput} shows that, at 64K context length, \system achieves \adddata{42} tokens/s, surpassing ShadowKV, SpeCache, and vLLM by \adddata{1.14$\times$}, \adddata{9.02$\times$}, and \adddata{3.65$\times$}, respectively. As the context length increases to 128K and 256K, vLLM consistently encounters out-of-memory (OOM) errors, while \system maintains high throughput and exceeds ShadowKV by \adddata{2.62$\times$} and \adddata{1.32$\times$}, respectively. SpeCache performs substantially worse, achieving \adddata{8.63$\times$} lower throughput than \system at 128K and failing with OOM at 256K. These results demonstrate the effectiveness of \system in sustaining high decoding throughput under increasingly memory-intensive long-context workloads. Meanwhile, we shows that the serving latencies (TPOT and TTFT) are not affected in \cref{sec: serving latencies}.

\begin{figure*}
    \centering
    \includegraphics[width=0.9\linewidth]{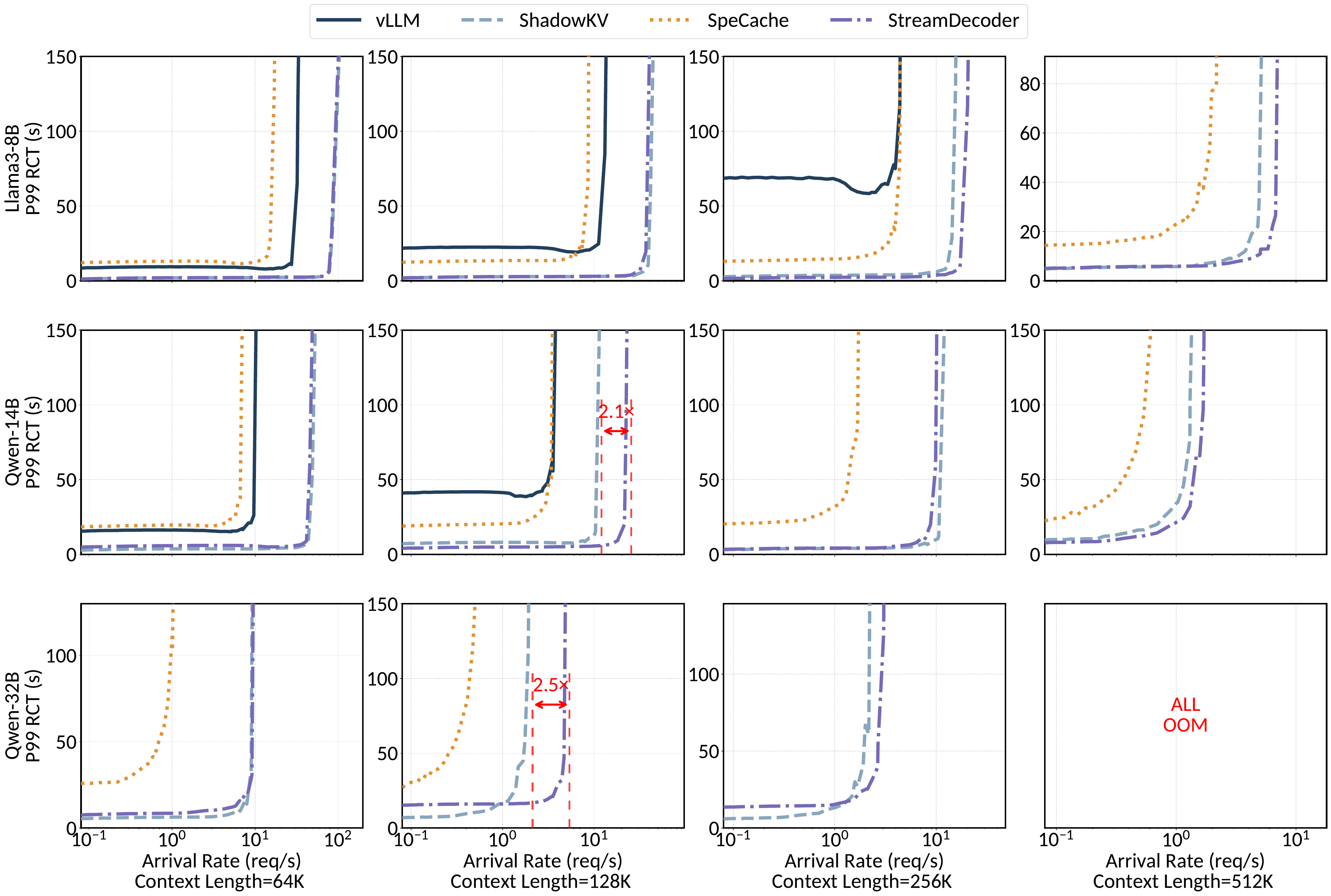}
    \vspace{-0.3cm}
        \caption{P99 request completion time across different arrival rates for the 32B model serving.}
    \vspace{-0.4cm}
    \label{fig:eval/RCT}
\end{figure*}

\noindent\textbf{Serving request concurrency.} We then evaluate the P99 request completion time (RCT) and its corresponding serving request rate at online serving cases. Following the settings in previous works~\cite{DBLP:conf/osdi/ZhongLCHZL0024,DBLP:conf/osdi/AgrawalKPMKGTR24}, the request arrivals are modeled as a Poisson process. Figure \ref{fig:eval/RCT} shows that when we set the P99 RCT goal at \addData{60s}, \system can tolerate up to \addData{$2.51 \times$} request arrival rate than shadowkv. For example,when serving a 32B model at a 128K context length when the request arrival rate is \addData{4.8} request per second, all requests can be completed in \addData{50} seconds for \system, where shadowkv's P99 request completion time increases exposively. And ShadowKV fails to satisfy the P99 RCT requirement once the request arrival rate exceeds \addData{1.92}. The reason is that \system benefits from a higher batch size running at a step, thereby reaching higher concurrency to serve requests.

\begin{figure}
    \centering
    \includegraphics[width=0.9\linewidth]{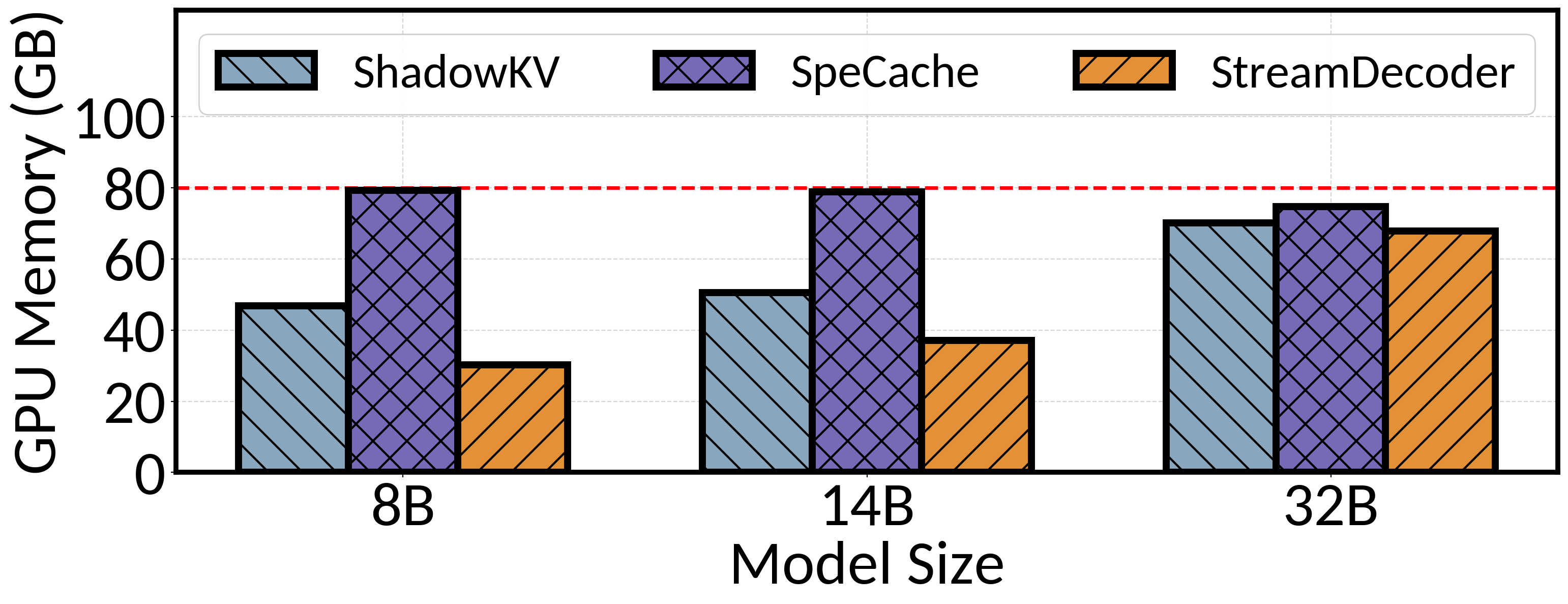}
    \vspace{-0.3cm}
    \caption{Memory footprint over the same batch sizes (12, 6 and 4 for 8B, 14B and 32B models respectively).}
    \vspace{-0.4cm}
    \label{fig:eval/Memory}
\end{figure}

\noindent\textbf{Memory footprint reduction.} To demonstrate \system efficiently reduces the serving memory usage, we hereby trace the memory footprint when serving different models. Figure \ref{fig:eval/Memory} shows that \system can reduce up to \addData{$36.4\%$} and \addData{$62.0\%$} of memory consumption compared to ShadowKV and SpeCache. For example, when serving batch size at \addData{12}, which is the biggest batch that SpeCache, the 8B model serving memory consumption of ShadowKV is \addData{47.5}GB and memory consumption of SpeCache is \addData{79}GB . In contrast, \system takes only \addData{30GB} memory. When serving 32B model with a batch size of \addData{4}, at which ShadowKV and SpeCache reach the limit of GPU HBM, our system can still save over\addData{13.8\%} memory consumption, which indicates \system can potentially serve larger batch or larger model on demand.

\begin{figure}
    \centering
    \includegraphics[width=0.9\linewidth]
    {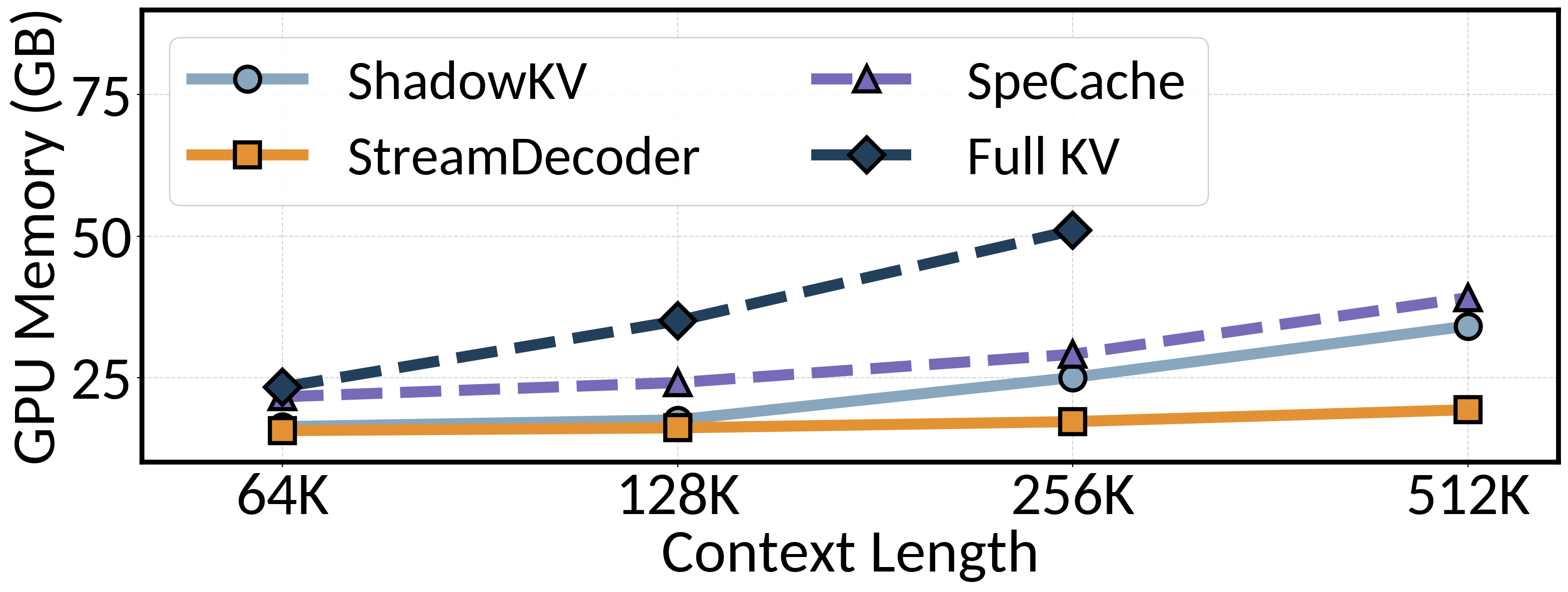}
    \vspace{-0.3cm}
    \caption{Memory footprint across different context lengths.}
    \vspace{-0.4cm}
    \label{fig:eval/scalability}
\end{figure}

\noindent\textbf{Scalability to context length} We further evaluate the GPU memory footprint scalability over context length, scaling from \adddata{128K} up to \adddata{512K} tokens. Figure \ref{fig:eval/scalability} shows that as the context length exponentially grows, the memory consumption of \system exhibits relatively slow growth compared to the linear explosion seen in baseline systems. For example, when extending the context length from \adddata{64K} to \adddata{512K}, the memory footprint of \system only increases by \adddata{3.7} GB, while ShadowKV and vLLM face OOM errors or require \adddata{17.8} GB of additional memory. 

The reason is that \system removes \residency, meanwhile, using layer-scoped KV memory management decouples the sparse KV working set from the total context length, significantly reducing the memory pressure.

\begin{figure}
    \centering
    \includegraphics[width=0.9\linewidth]{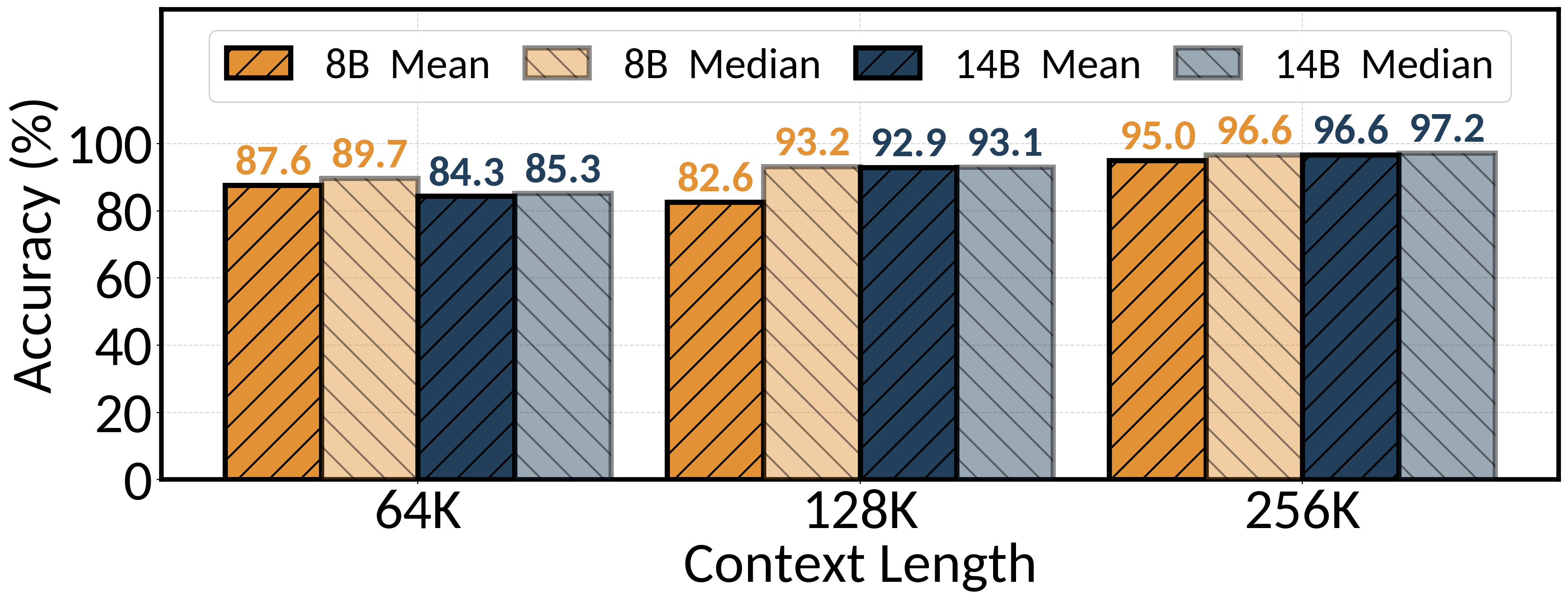}
    \vspace{-0.3cm}
    \caption{KV retrieval prediction accuracy across models.}
    \vspace{-0.4cm}
    \label{fig:eval/index}
\end{figure}

\subsection{Accuracy Analysis}
\label{sec: Accuracy Analysis}
\noindent \textbf{KV cache prediction accuracy.} Then we evaluated the KV cache prediction accuracy by tracking the overlap ratio between the predicted sparse indices and the ground-truth indices during continuous decoding across two representative datasets, including NIAH-1 and NIAH-2. Figure \ref{fig:eval/index} shows that \system consistently achieves a highly stable and accurate prediction hit rate regardless of the dataset variations. For example, on the \addData{Ruler} dataset with a 128K context length, the accuracy stays above \addData{82.6}\%, meaning that the missed KV entries only account for a marginal \addData{17.4}\% of the total required tokens. These results demonstrate that \system can accurately capture the essential KV pairs ahead of time, proving that the significant KV dependencies between two consecutive tokens are highly coincident. 

\noindent \textbf{End-to-end serving accuracy.} To show that \system can increase the serving throughput without impairing serving accuracy, we evaluate the end-to-end downstream task model accuracy on datasets including NIAH-1, NIAH-2, NIAH-3 and QA-2 of RULER. Table \ref{tab:niah_accuracy} shows that \system maintains strictly lossless quality compared to ShadowKV, in most of the datasets. This is reasonable since \system handles the missed KV entries via our prioritized scheduling, thereby showing the identical behaviors compared to common sparse KV cache serving systems. In contrast, ShadowKV can suffer from the accuracy degradation due to the low-rank key reconstruction, while \system can relieve this disadvantage as shown in the accuracy of the NIAH-3 dataset.

\begin{table}
\centering
\caption{End-to-end NIAH accuracy comparison between \textsc{ShadowKV} and \system(ours).
We use Llama-3-8B-Instruct-Gradient-1048k with a sparse context length of 1024.
Exact-match accuracy of the first generated token against the needle value.}
\label{tab:niah_accuracy}
\begin{tabular}{l cccc}
\toprule
\multirow{2}{*}{Dataset} & \multicolumn{2}{c}{64K Context} & \multicolumn{2}{c}{128K Context} \\
\cmidrule(lr){2-3} \cmidrule(lr){4-5}
 & ShadowKV & Ours & ShadowKV & Ours \\
\midrule
NIAH-1 & 1.00 & \textbf{1.00} & 1.00 & \textbf{1.00} \\
NIAH-2 & 1.00 & \textbf{1.00} & 1.00 & \textbf{1.00} \\
NIAH-3 & 0.82& \textbf{1.00} & 0.24& \textbf{1.00} \\
QA-2          & 0.53& 0.58& 0.38& 0.43\\
\midrule
\textbf{Avg} & 0.94& \textbf{1.00} & 0.74& \textbf{1.00} \\
\bottomrule
\end{tabular}
\end{table}

\subsection{Modular Study}
\label{sec: Modular Study}
\begin{figure}[!tb]
    \centering
    \subfigure[Attention layers.]{
        \label{fig: eval/flashattn}
        \hspace{-0.28cm}
        \includegraphics[width=0.48\linewidth]{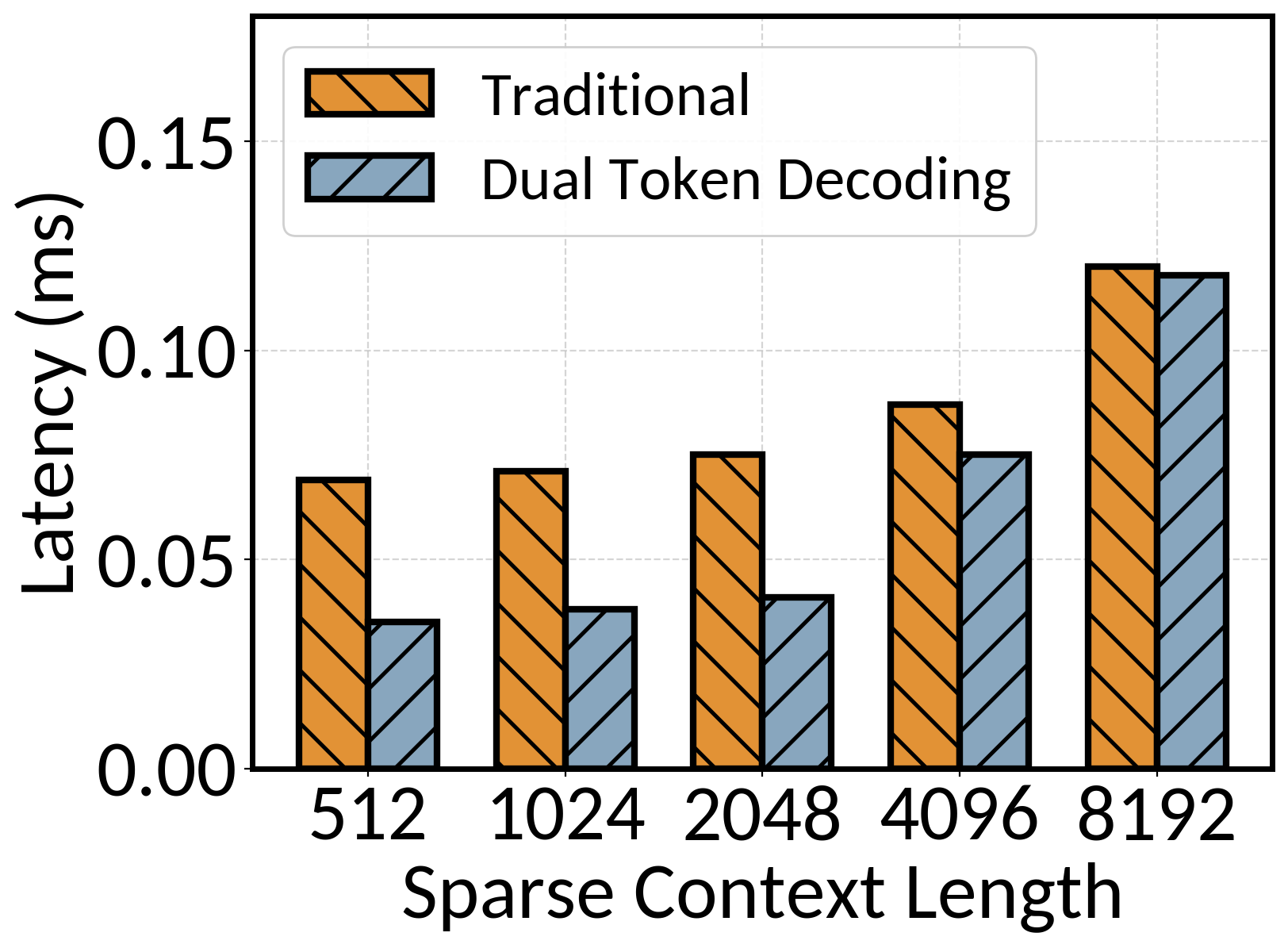}
    }
   \hspace{-0.2cm}
    \subfigure[Linear layers.]{
        \label{fig: eval/gemm}
        \includegraphics[width=0.48\linewidth]{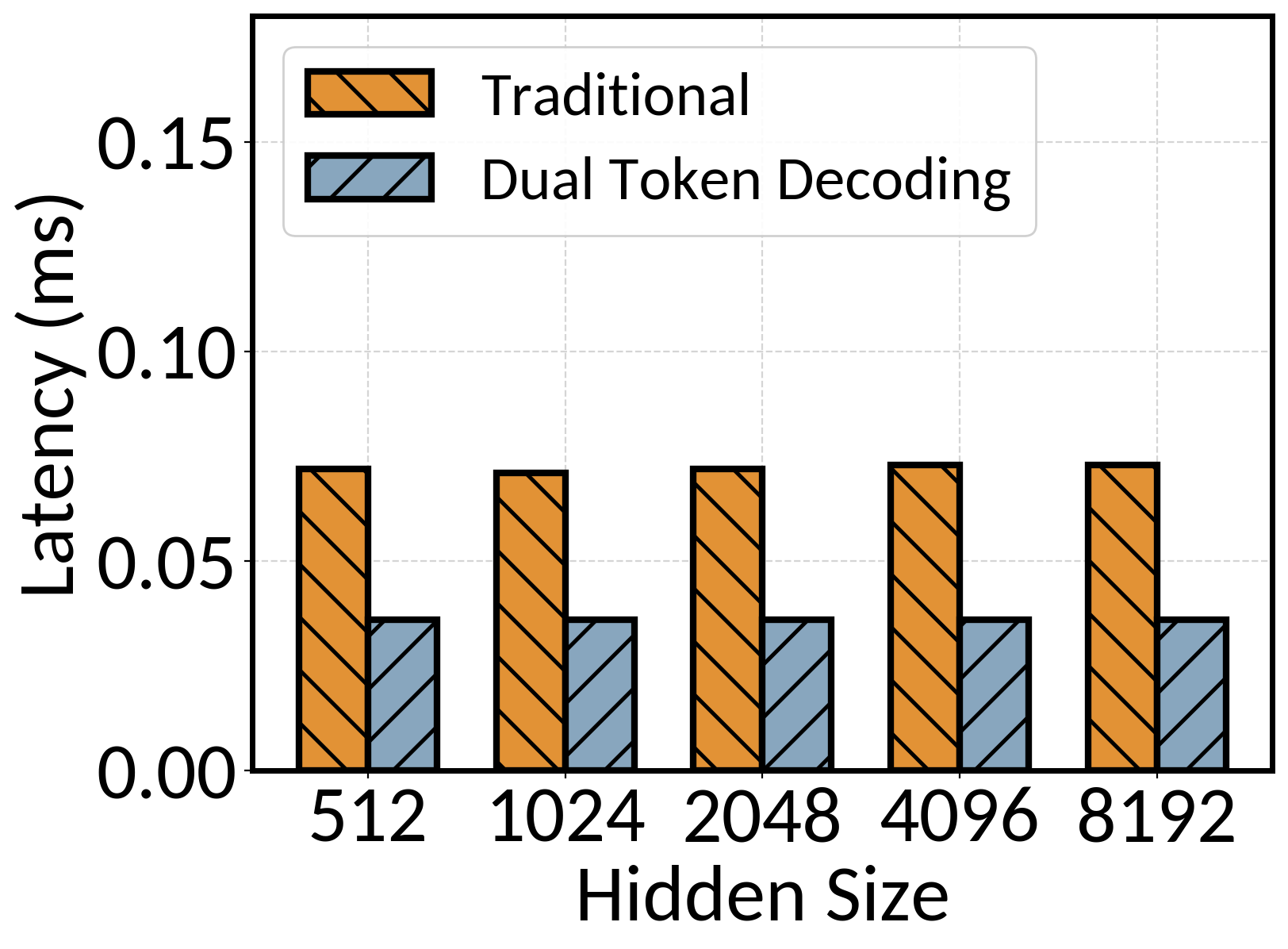}
    }
    \vspace{-0.3cm}
    \caption{Layer computation efficiency with \system dual-token decoding.}
    \label{fig: eval/dual-token decoding}
    \vspace{-0.4cm}
\end{figure}

\noindent \textbf{Dual-token decoding efficiency.} We first evaluate the effectiveness of our dual-token execution optimization. The baseline approach is a traditional and naive execution design that runs an extra decoding pass to generate the \spectoken. As shown in Figure \ref{fig: eval/dual-token decoding}, by stacking the output and speculative tokens into a single batch dimension, \system drastically reduces both the computational latencies of attention and linear layers. For example, computing two tokens via our Co-batched attention takes only \addData{0.035} ms, saving \addData{49}\% of the time compared to the \addData{0.069} ms required by consecutively launching two separate passes.

\begin{figure}[!tb]
    \centering
    \subfigure[8B model serving.]{
        \label{fig: eval/priority1}
        \hspace{-0.28cm}
        \includegraphics[width=0.48\linewidth]{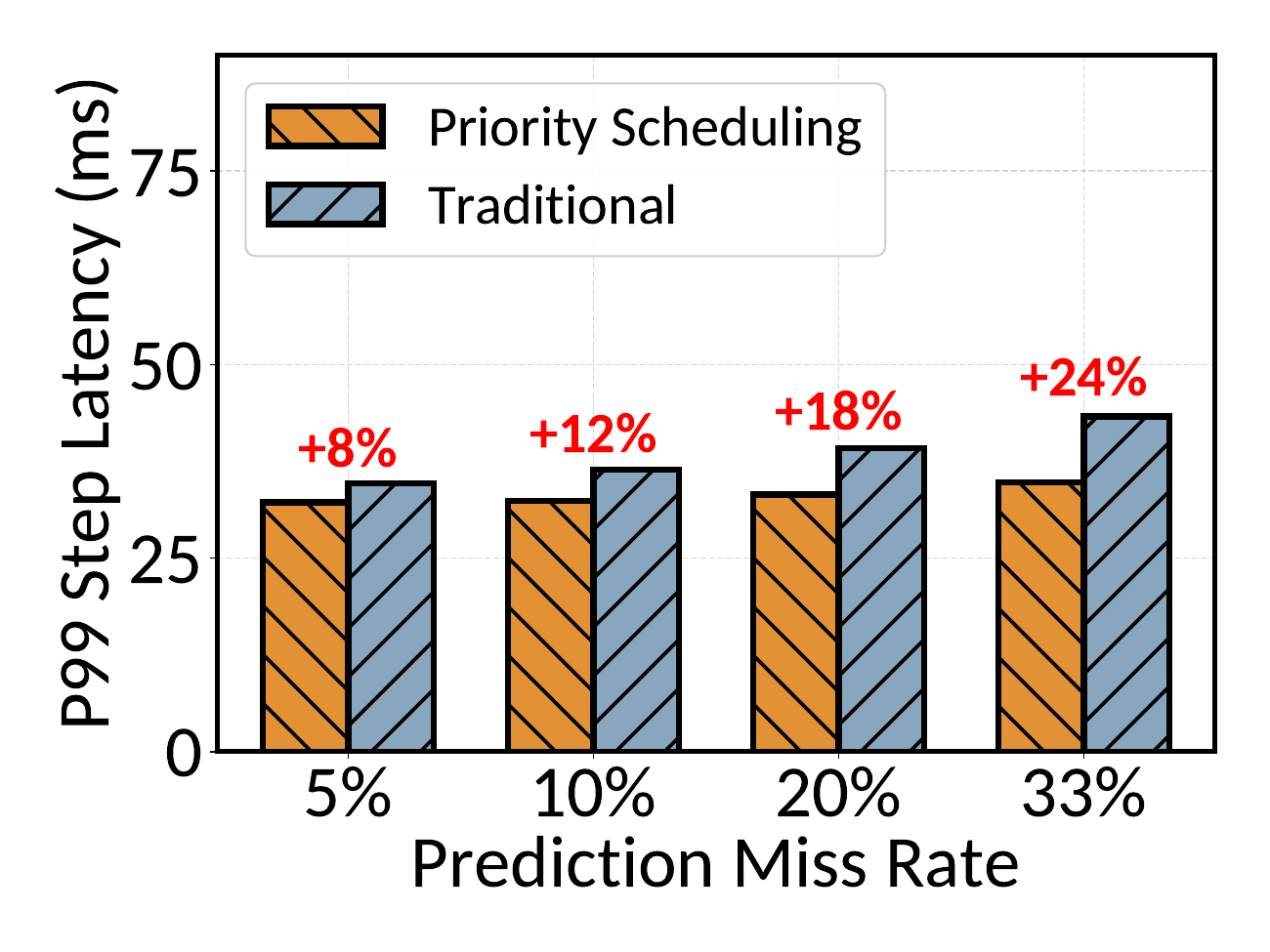}
    }
   \hspace{-0.2cm}
    \subfigure[14B model serving.]{
        \label{fig: eval/priority2}
        \includegraphics[width=0.48\linewidth]{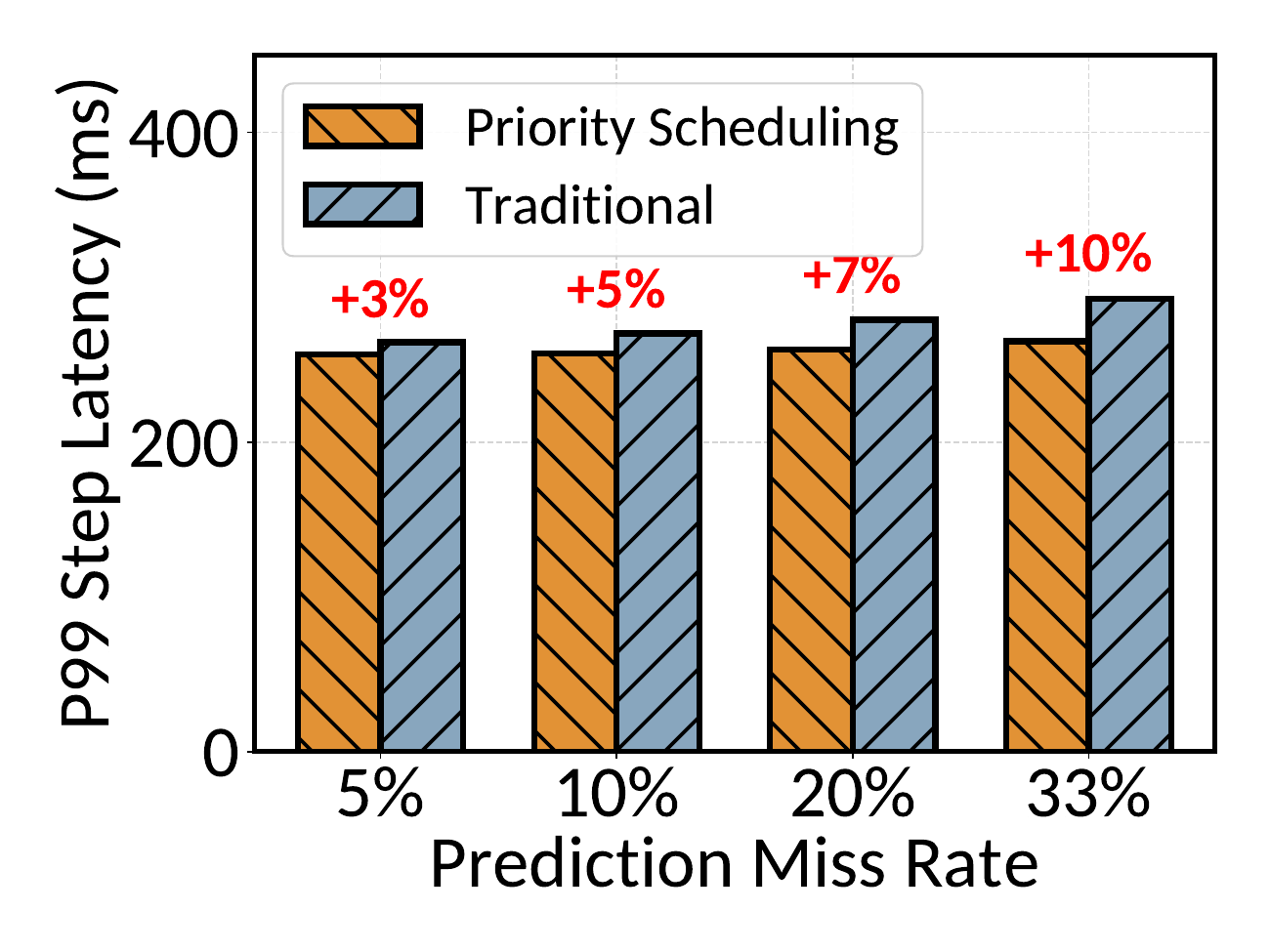}
    }
    \vspace{-0.3cm}
    \caption{Per-step latency with \system priority scheduling for missed KV entries.}
    \label{fig: eval/priority scheduling}
    \vspace{-0.4cm}
\end{figure}

\noindent\textbf{Priority scheduling.} Then we evaluate the impact of our two-stream priority scheduling by comparing it against a naive scheduling baseline where all CPU-to-GPU transfer tasks queue sequentially without priorities. Figure \ref{fig: eval/priority scheduling},\ref{fig: eval/priority2} shows that \system's priority scheduling successfully eliminates the severe pipeline stalls caused by prediction errors in the critical path. For example, under a prediction miss rate of \addData{33}\% on a 8B model, our scheduling restricts the pipeline blocking time to merely \addData{34.8} ms, whereas the naive approach suffers up to \addData{43.3} ms of delay. The prioritized schedule ensures that occasional mispredictions do not disrupt the continuous computation flow, maintaining a highly stable token generation speed over the naive baseline. 

\begin{figure}
    \centering
    \includegraphics[width=0.9\linewidth]{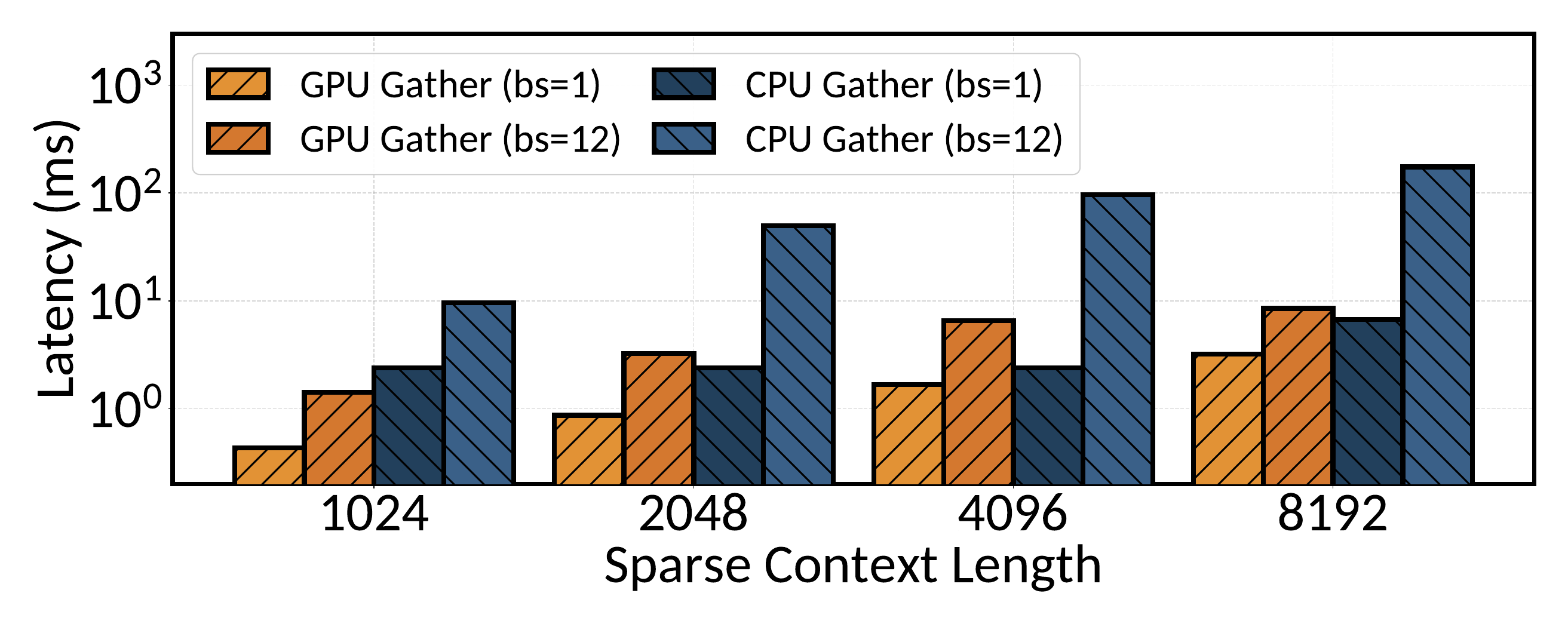}
    \vspace{-0.3cm}
    \caption{KV retrieval latency using \system GPU gathering.}
    \label{fig:eval/gather}
    \vspace{-0.5cm}
\end{figure}

\noindent\textbf{GPU gather copy latencies.} \system utilizes a GPU-side KV entry gather copy kernel to rapidly transfer the discrete CPU KV entries. We hereby evaluate this design against traditional CPU-side gather transfers. Figure \ref{fig:eval/gather} shows that offloading the KV gather copy operations to the GPU dramatically reduces both the I/O bandwidth consumption and the host-device transmission latency. For example, our GPU gather kernel lowers the total KV transmission overhead to \addData{3.25} ms, achieving a \addData{93.4}\% reduction compared to the \addData{49.5} ms bottleneck caused by frequent random reads in the CPU Gather approach, when serving a sparse context length of \addData{2048}. 
The significant improvement stems from that \system performs asynchronous, contiguous block prefetching over PCIe and leverages the massive parallelism of GPU kernels to execute a monolithic memory reorder only after the ground-truth tokens are confirmed.

\section{Related Work}
\label{sec: related work}

\noindent\textbf{Dynamic sparse KV cache techniques.}
A substantial body of work has explored dynamic sparse attention for long-context inference. DiffKV \cite{DBLP:conf/sosp/ZhangHZLC25}, HeadKV \cite{DBLP:conf/iclr/FuCAX0X25}, and PyramidKV \cite{DBLP:conf/acl/YangHGHZ024} leverage sparsity in numerical precision, attention heads, and transformer layers, respectively. 
StreamingLLM \cite{xiao2023streamingllm} keeps initial tokens in a sequence referred to as ``attention sinks'', capturing an disproportionately large share of attention scores. DuoAttention \cite{DBLP:conf/iclr/XiaoTZGYTF025} uses optimization-based profiling to select critical retrieval heads and preserves full context only for retrieval heads.

KV offloading mitigates GPU memory pressure by transferring the KV cache to host memory and fetching only the required entries on demand. ShadowKV \cite{lin2024shadowkv} maintains a low-rank pre-RoPE Key cache on the GPU using SVD while offloading the value cache to host memory. Upon model inference, it identifies the most relevant Key cache chunks, and overlaps sparse value retrieval with key reconstruction to hide transfer latency. SpeCache \cite{chen2024specache} employs a draft model and exploits the similarity between the sparse KV indices selected by speculative tokens and those required by the actual output tokens, enabling proactive KV prefetching. SpeContext \cite{DBLP:conf/asplos/XuPWZY0026} uses a distilled language model-based retrieval head to choose useful KV pairs, and overlaps KV-cache prefetching with LLM computation. In contrast, our approach reduces the GPU-resident sparse-KV auxiliary state to a token-independent and layer-independent footprint.

\noindent\textbf{Long context serving systems.} 
Systems for long-context LLM serving have evolved rapidly in recent years. One of the earliest efforts, LoongServe \cite{DBLP:conf/sosp/WuLZ0L024}, employs sequence parallelism to reduce the memory footprint of KV caches for extremely long contexts. To accommodate highly dynamic inference workloads, LoongServe further introduces elastic sequence parallelism, which adaptively determines the degree of parallelism for each batch.
MemServe \cite{DBLP:journals/corr/abs-2406-17565} and DualPath \cite{DBLP:journals/corr/abs-2602-21548} utilize remote KV-cache pools to extend the storage capacity available for long-context serving. MemServe introduces a distributed memory pool that manages KV caches across serving instances, significantly increasing the effective KV-cache capacity. DualPath focuses on efficient retrieval of KV caches from remote persistent storage and addresses the communication bottlenecks associated with large-scale KV-cache transfers. 
Our work focused on long context serving via sparse KV retrieval, which can be benifit from the KV Pool and sequence parallel techniques and furthermore improve the performance with these orthogonal works.

\section{Conclusion}
\label{sec: conclusion}

This paper presents \system, a lightweight long-context LLM system that accelerates sparse KV-cache retrieval with KV prefetching. It exploits the strong \emph{predictability} of sparse KV indices across consecutive decoding tokens to optimize future retrievals. \system designs a dual-token decoding pipeline, layer-aware KV transfer scheduling, prioritized recovery for missing KV entries, and adaptive key reconstruction to overlap communication with execution and reduce critical-path transfers. Experiments show that \system improves decoding throughput, lowers GPU memory use, and maintains accuracy on long-context workloads. 
\label{lastpage}
\end{sloppypar}

\bibliographystyle{ACM-Reference-Format}
\bibliography{sample}

@article{DBLP:journals/corr/abs-2406-17565,
  author       = {Cunchen Hu and
                  Heyang Huang and
                  Junhao Hu and
                  Jiang Xu and
                  Xusheng Chen and
                  Tao Xie and
                  Chenxi Wang and
                  Sa Wang and
                  Yungang Bao and
                  Ninghui Sun and
                  Yizhou Shan},
  title        = {MemServe: Context Caching for Disaggregated {LLM} Serving with Elastic
                  Memory Pool},
  journal      = {CoRR},
  volume       = {abs/2406.17565},
  year         = {2024},
  url          = {https://doi.org/10.48550/arXiv.2406.17565},
  doi          = {10.48550/ARXIV.2406.17565},
  eprinttype   = {arXiv},
  eprint       = {2406.17565},
  bibsource    = {dblp computer science bibliography, https://dblp.org}
}

@inproceedings{DBLP:conf/icml/ShangZWZL0C025,
  author       = {Ning Shang and
                  Li Lyna Zhang and
                  Siyuan Wang and
                  Gaokai Zhang and
                  Gilsinia Lopez and
                  Fan Yang and
                  Weizhu Chen and
                  Mao Yang},
  editor       = {Aarti Singh and
                  Maryam Fazel and
                  Daniel Hsu and
                  Simon Lacoste{-}Julien and
                  Felix Berkenkamp and
                  Tegan Maharaj and
                  Kiri Wagstaff and
                  Jerry Zhu},
  title        = {LongRoPE2: Near-Lossless {LLM} Context Window Scaling},
  booktitle    = {Forty-second International Conference on Machine Learning, {ICML}
                  2025, Vancouver, BC, Canada, July 13-19, 2025},
  series       = {Proceedings of Machine Learning Research},
  publisher    = {{PMLR} / OpenReview.net},
  year         = {2025},
  url          = {https://proceedings.mlr.press/v267/shang25a.html},
  bibsource    = {dblp computer science bibliography, https://dblp.org}
}

@inproceedings{DBLP:conf/icml/DingZZXSX0Y24,
  author       = {Yiran Ding and
                  Li Lyna Zhang and
                  Chengruidong Zhang and
                  Yuanyuan Xu and
                  Ning Shang and
                  Jiahang Xu and
                  Fan Yang and
                  Mao Yang},
  editor       = {Ruslan Salakhutdinov and
                  Zico Kolter and
                  Katherine A. Heller and
                  Adrian Weller and
                  Nuria Oliver and
                  Jonathan Scarlett and
                  Felix Berkenkamp},
  title        = {LongRoPE: Extending {LLM} Context Window Beyond 2 Million Tokens},
  booktitle    = {Forty-first International Conference on Machine Learning, {ICML} 2024,
                  Vienna, Austria, July 21-27, 2024},
  series       = {Proceedings of Machine Learning Research},
  pages        = {11091--11104},
  publisher    = {{PMLR} / OpenReview.net},
  year         = {2024},
  url          = {https://proceedings.mlr.press/v235/ding24i.html},
  bibsource    = {dblp computer science bibliography, https://dblp.org}
}

@inproceedings{DBLP:conf/emnlp/0019ZHLHDGZ024,
  author       = {Jun Zhao and
                  Can Zu and
                  Xu Hao and
                  Yi Lu and
                  Wei He and
                  Yiwen Ding and
                  Tao Gui and
                  Qi Zhang and
                  Xuanjing Huang},
  editor       = {Yaser Al{-}Onaizan and
                  Mohit Bansal and
                  Yun{-}Nung Chen},
  title        = {{LONGAGENT:} Achieving Question Answering for 128k-Token-Long Documents
                  through Multi-Agent Collaboration},
  booktitle    = {Proceedings of the 2024 Conference on Empirical Methods in Natural
                  Language Processing, {EMNLP} 2024, Miami, FL, USA, November 12-16,
                  2024},
  pages        = {16310--16324},
  publisher    = {Association for Computational Linguistics},
  year         = {2024},
  url          = {https://doi.org/10.18653/v1/2024.emnlp-main.912},
  doi          = {10.18653/V1/2024.EMNLP-MAIN.912},
  bibsource    = {dblp computer science bibliography, https://dblp.org}
}

@article{DBLP:journals/corr/abs-2602-21548,
  author       = {Yongtong Wu and
                  Shaoyuan Chen and
                  Yinmin Zhong and
                  Rilin Huang and
                  Yixuan Tan and
                  Wentao Zhang and
                  Liyue Zhang and
                  Shangyan Zhou and
                  Yuxuan Liu and
                  Shunfeng Zhou and
                  Mingxing Zhang and
                  Xin Jin and
                  Panpan Huang},
  title        = {DualPath: Breaking the Storage Bandwidth Bottleneck in Agentic {LLM}
                  Inference},
  journal      = {CoRR},
  volume       = {abs/2602.21548},
  year         = {2026},
  url          = {https://doi.org/10.48550/arXiv.2602.21548},
  doi          = {10.48550/ARXIV.2602.21548},
  eprinttype   = {arXiv},
  eprint       = {2602.21548},
  bibsource    = {dblp computer science bibliography, https://dblp.org}
}

@inproceedings{DBLP:conf/sosp/WuLZ0L024,
  author       = {Bingyang Wu and
                  Shengyu Liu and
                  Yinmin Zhong and
                  Peng Sun and
                  Xuanzhe Liu and
                  Xin Jin},
  editor       = {Emmett Witchel and
                  Christopher J. Rossbach and
                  Andrea C. Arpaci{-}Dusseau and
                  Kimberly Keeton},
  title        = {LoongServe: Efficiently Serving Long-Context Large Language Models
                  with Elastic Sequence Parallelism},
  booktitle    = {Proceedings of the {ACM} {SIGOPS} 30th Symposium on Operating Systems
                  Principles, {SOSP} 2024, Austin, TX, USA, November 4-6, 2024},
  pages        = {640--654},
  publisher    = {{ACM}},
  year         = {2024},
  url          = {https://doi.org/10.1145/3694715.3695948},
  doi          = {10.1145/3694715.3695948},
  bibsource    = {dblp computer science bibliography, https://dblp.org}
}

@inproceedings{kwon2023vllm,
  title     = {Efficient Memory Management for Large Language Model Serving with PagedAttention},
  author    = {Kwon, Woosuk and Li, Zhuohan and Zhuang, Siyuan and Sheng, Ying and Zheng, Lianmin and Yu, Cody Hao and Gonzalez, Joseph E. and Zhang, Hao and Stoica, Ion},
  booktitle = {Proceedings of the 29th ACM Symposium on Operating Systems Principles (SOSP)},
  year      = {2023},
  pages     = {611--626},
  doi       = {10.1145/3600006.3613165},
  url       = {https://dl.acm.org/doi/10.1145/3600006.3613165}
}

@inproceedings{xiao2023streamingllm,
  title     = {StreamingLLM: Efficient Streaming Inference of Large Language Models with Attention Sinks},
  author    = {Xiao, Guangxuan and Tian, Yaohui and Chen, Beidi and others},
  booktitle = {International Conference on Learning Representations (ICLR)},
  year      = {2024},
  url       = {https://iclr.cc/media/iclr-2024/Slides/18794.pdf},
  note      = {Slides.}
}

@inproceedings{zhang2023h2o,
  title     = {H2O: Heavy-Hitter Oracle for Efficient Generative Inference of Large Language Models},
  author    = {Zhang, Zhenyu and Sheng, Ying and Zhou, Tianyi and Chen, Tianlong and Zheng, Lianmin and Cai, Ruisi and Song, Zhao and Tian, Yuandong and R{\'e}, Christopher and Barrett, Clark W. and Wang, Zhangyang and Chen, Beidi},
  booktitle = {Advances in Neural Information Processing Systems (NeurIPS)},
  year      = {2023},
  url       = {https://dl.acm.org/doi/10.5555/3666122.3667628}
}

@inproceedings{lin2024shadowkv,
  title     = {ShadowKV: KV Cache in Shadows for High-Throughput Long-Context LLM Inference},
  author    = {Sun, Hanshi and Chang, Li-Wen and Bao, Wenlei and Zheng, Size and Zheng, Ningxin and Liu, Xin and Dong, Harry and Chi, Yuejie and Chen, Beidi},
  booktitle = {International Conference on Machine Learning (ICML)},
  year      = {2025},
  url       = {https://icml.cc/virtual/2025/poster/44053}
}

@inproceedings{chen2024specache,
  title     = {SpeCache: Speculative Key-Value Caching for Efficient Generation of LLMs},
  author    = {Jie, Shibo and Tang, Yehui and Han, Kai and Deng, Zhi-Hong and Han, Jing},
  booktitle = {International Conference on Machine Learning (ICML)},
  year      = {2025},
  url       = {https://icml.cc/virtual/2025/poster/45386}
}

@article{leviathan2023speculative,
  title   = {Fast Inference from Transformers via Speculative Decoding},
  author  = {Leviathan, Yaniv and Kalman, Matan and Matias, Yossi},
  journal = {International Conference on Machine Learning (ICML)},
  year    = {2023}
}

@article{schuster2023lookahead,
  title   = {Accelerating Large Language Model Decoding with Speculative Sampling},
  author  = {Schuster, Tal and others},
  journal = {arXiv preprint arXiv:2302.01318},
  year    = {2023}
}

@inproceedings{DBLP:conf/isca/XuPZCLLWD25,
  author       = {Jiaming Xu and
                  Jiayi Pan and
                  Yongkang Zhou and
                  Siming Chen and
                  Jinhao Li and
                  Yaoxiu Lian and
                  Junyi Wu and
                  Guohao Dai},
  title        = {SpecEE: Accelerating Large Language Model Inference with Speculative
                  Early Exiting},
  booktitle    = {Proceedings of the 52nd Annual International Symposium on Computer
                  Architecture, {ISCA} 2025, Tokyo, Japan, June 21-25, 2025},
  pages        = {467--481},
  publisher    = {{ACM}},
  year         = {2025},
  url          = {https://doi.org/10.1145/3695053.3730996},
  doi          = {10.1145/3695053.3730996},
  bibsource    = {dblp computer science bibliography, https://dblp.org}
}

@inproceedings{DBLP:conf/icml/GloeckleIRLS24,
  author       = {Fabian Gloeckle and
                  Badr Youbi Idrissi and
                  Baptiste Rozi{\`{e}}re and
                  David Lopez{-}Paz and
                  Gabriel Synnaeve},
  editor       = {Ruslan Salakhutdinov and
                  Zico Kolter and
                  Katherine A. Heller and
                  Adrian Weller and
                  Nuria Oliver and
                  Jonathan Scarlett and
                  Felix Berkenkamp},
  title        = {Better {\&} Faster Large Language Models via Multi-token Prediction},
  booktitle    = {Forty-first International Conference on Machine Learning, {ICML} 2024,
                  Vienna, Austria, July 21-27, 2024},
  series       = {Proceedings of Machine Learning Research},
  pages        = {15706--15734},
  publisher    = {{PMLR} / OpenReview.net},
  year         = {2024},
  url          = {https://proceedings.mlr.press/v235/gloeckle24a.html},
  bibsource    = {dblp computer science bibliography, https://dblp.org}
}

@inproceedings{DBLP:conf/acl/ElhoushiSLHWL0A24,
  author       = {Mostafa Elhoushi and
                  Akshat Shrivastava and
                  Diana Liskovich and
                  Basil Hosmer and
                  Bram Wasti and
                  Liangzhen Lai and
                  Anas Mahmoud and
                  Bilge Acun and
                  Saurabh Agarwal and
                  Ahmed Roman and
                  Ahmed A. Aly and
                  Beidi Chen and
                  Carole{-}Jean Wu},
  editor       = {Lun{-}Wei Ku and
                  Andre Martins and
                  Vivek Srikumar},
  title        = {LayerSkip: Enabling Early Exit Inference and Self-Speculative Decoding},
  booktitle    = {Proceedings of the 62nd Annual Meeting of the Association for Computational
                  Linguistics (Volume 1: Long Papers), {ACL} 2024, Bangkok, Thailand,
                  August 11-16, 2024},
  pages        = {12622--12642},
  publisher    = {Association for Computational Linguistics},
  year         = {2024},
  url          = {https://doi.org/10.18653/v1/2024.acl-long.681},
  doi          = {10.18653/V1/2024.ACL-LONG.681},
  bibsource    = {dblp computer science bibliography, https://dblp.org}
}

@inproceedings{jain2023orca,
  title     = {Orca: A Distributed Serving System for Transformer-Based Generative Models},
  author    = {Yu, Gyeong-In and Jeong, Joo Seong and Kim, Geon-Woo and Kim, Soojeong and Chun, Byung-Gon},
  booktitle = {16th {USENIX} Symposium on Operating Systems Design and Implementation (OSDI)},
  pages     = {521--538},
  year      = {2022},
  url       = {https://dblp.org/rec/conf/osdi/YuJKKC22}
}

@inproceedings{DBLP:conf/osdi/AgrawalKPMKGTR24,
  author       = {Amey Agrawal and
                  Nitin Kedia and
                  Ashish Panwar and
                  Jayashree Mohan and
                  Nipun Kwatra and
                  Bhargav S. Gulavani and
                  Alexey Tumanov and
                  Ramachandran Ramjee},
  editor       = {Ada Gavrilovska and
                  Douglas B. Terry},
  title        = {Taming Throughput-Latency Tradeoff in {LLM} Inference with Sarathi-Serve},
  booktitle    = {18th {USENIX} Symposium on Operating Systems Design and Implementation,
                  {OSDI} 2024, Santa Clara, CA, USA, July 10-12, 2024},
  pages        = {117--134},
  publisher    = {{USENIX} Association},
  year         = {2024},
  url          = {https://www.usenix.org/conference/osdi24/presentation/agrawal},
  bibsource    = {dblp computer science bibliography, https://dblp.org}
}

@inproceedings{DBLP:conf/osdi/ZhongLCHZL0024,
	title = {{DistServe}: {Disaggregating} prefill and decoding for goodput-optimized large language model serving},
	url = {https://www.usenix.org/conference/osdi24/presentation/zhong-yinmin},
	booktitle = {18th {USENIX} symposium on operating systems design and implementation, {OSDI} 2024, santa clara, {CA}, {USA}, july 10-12, 2024},
	publisher = {USENIX Association},
	author = {Zhong, Yinmin and Liu, Shengyu and Chen, Junda and Hu, Jianbo and Zhu, Yibo and Liu, Xuanzhe and Jin, Xin and Zhang, Hao},
	editor = {Gavrilovska, Ada and Terry, Douglas B.},
	year = {2024},
	note = {tex.bibsource: dblp computer science bibliography, https://dblp.org
tex.timestamp: Tue, 11 Feb 2025 11:42:30 +0100},
	pages = {193--210},
}

@inproceedings{DBLP:conf/euromlsys/HeLA24,
  author       = {Yongjun He and
                  Yao Lu and
                  Gustavo Alonso},
  title        = {Deferred Continuous Batching in Resource-Efficient Large Language
                  Model Serving},
  booktitle    = {Proceedings of the 4th Workshop on Machine Learning and Systems, EuroMLSys
                  2024, Athens, Greece, 22 April 2024},
  pages        = {98--106},
  publisher    = {{ACM}},
  year         = {2024},
  url          = {https://doi.org/10.1145/3642970.3655835},
  doi          = {10.1145/3642970.3655835},
  bibsource    = {dblp computer science bibliography, https://dblp.org}
}

@inproceedings{DBLP:conf/acl/BaiTZ0WLCX0D0L25,
  author       = {Yushi Bai and
                  Shangqing Tu and
                  Jiajie Zhang and
                  Hao Peng and
                  Xiaozhi Wang and
                  Xin Lv and
                  Shulin Cao and
                  Jiazheng Xu and
                  Lei Hou and
                  Yuxiao Dong and
                  Jie Tang and
                  Juanzi Li},
  editor       = {Wanxiang Che and
                  Joyce Nabende and
                  Ekaterina Shutova and
                  Mohammad Taher Pilehvar},
  title        = {LongBench v2: Towards Deeper Understanding and Reasoning on Realistic
                  Long-context Multitasks},
  booktitle    = {Proceedings of the 63rd Annual Meeting of the Association for Computational
                  Linguistics (Volume 1: Long Papers), {ACL} 2025, Vienna, Austria,
                  July 27 - August 1, 2025},
  pages        = {3639--3664},
  publisher    = {Association for Computational Linguistics},
  year         = {2025},
  url          = {https://doi.org/10.18653/v1/2025.acl-long.183},
  doi          = {10.18653/V1/2025.ACL-LONG.183},
  bibsource    = {dblp computer science bibliography, https://dblp.org}
}

@article{DBLP:journals/corr/abs-2512-02556,
  author       = {DeepSeek{-}AI},
  title        = {DeepSeek-V3.2: Pushing the Frontier of Open Large Language Models},
  journal      = {CoRR},
  volume       = {abs/2512.02556},
  year         = {2025},
  url          = {https://doi.org/10.48550/arXiv.2512.02556},
  doi          = {10.48550/ARXIV.2512.02556},
  eprinttype   = {arXiv},
  eprint       = {2512.02556},
  bibsource    = {dblp computer science bibliography, https://dblp.org}
}

@article{DBLP:journals/corr/abs-1911-02150,
  author       = {Noam Shazeer},
  title        = {Fast Transformer Decoding: One Write-Head is All You Need},
  journal      = {CoRR},
  volume       = {abs/1911.02150},
  year         = {2019},
  url          = {http://arxiv.org/abs/1911.02150},
  eprinttype   = {arXiv},
  eprint       = {1911.02150},
  bibsource    = {dblp computer science bibliography, https://dblp.org}
}

@inproceedings{DBLP:conf/iclr/XiaoTZGYTF025,
  author       = {Guangxuan Xiao and
                  Jiaming Tang and
                  Jingwei Zuo and
                  Junxian Guo and
                  Shang Yang and
                  Haotian Tang and
                  Yao Fu and
                  Song Han},
  title        = {DuoAttention: Efficient Long-Context {LLM} Inference with Retrieval
                  and Streaming Heads},
  booktitle    = {The Thirteenth International Conference on Learning Representations,
                  {ICLR} 2025, Singapore, April 24-28, 2025},
  publisher    = {OpenReview.net},
  year         = {2025},
  url          = {https://openreview.net/forum?id=cFu7ze7xUm},
  bibsource    = {dblp computer science bibliography, https://dblp.org}
}

@inproceedings{DBLP:conf/sigcomm/LiuLCRHZDY0AMHH24,
  author       = {Yuhan Liu and
                  Hanchen Li and
                  Yihua Cheng and
                  Siddhant Ray and
                  Yuyang Huang and
                  Qizheng Zhang and
                  Kuntai Du and
                  Jiayi Yao and
                  Shan Lu and
                  Ganesh Ananthanarayanan and
                  Michael Maire and
                  Henry Hoffmann and
                  Ari Holtzman and
                  Junchen Jiang},
  title        = {CacheGen: {KV} Cache Compression and Streaming for Fast Large Language
                  Model Serving},
  booktitle    = {Proceedings of the {ACM} {SIGCOMM} 2024 Conference, {ACM} {SIGCOMM}
                  2024, Sydney, NSW, Australia, August 4-8, 2024},
  pages        = {38--56},
  publisher    = {{ACM}},
  year         = {2024},
  url          = {https://doi.org/10.1145/3651890.3672274},
  doi          = {10.1145/3651890.3672274},
  bibsource    = {dblp computer science bibliography, https://dblp.org}
}

@article{DBLP:journals/corr/abs-2403-11421,
  author       = {Jiaao He and
                  Jidong Zhai},
  title        = {FastDecode: High-Throughput GPU-Efficient {LLM} Serving using Heterogeneous
                  Pipelines},
  journal      = {CoRR},
  volume       = {abs/2403.11421},
  year         = {2024},
  url          = {https://doi.org/10.48550/arXiv.2403.11421},
  doi          = {10.48550/ARXIV.2403.11421},
  eprinttype   = {arXiv},
  eprint       = {2403.11421},
  bibsource    = {dblp computer science bibliography, https://dblp.org}
}

@inproceedings{DBLP:conf/eurosys/GaoCS25,
  author       = {Shiwei Gao and
                  Youmin Chen and
                  Jiwu Shu},
  title        = {Fast State Restoration in {LLM} Serving with HCache},
  booktitle    = {Proceedings of the Twentieth European Conference on Computer Systems,
                  EuroSys 2025, Rotterdam, The Netherlands, 30 March 2025 - 3 April
                  2025},
  pages        = {128--143},
  publisher    = {{ACM}},
  year         = {2025},
  url          = {https://doi.org/10.1145/3689031.3696072},
  doi          = {10.1145/3689031.3696072},
  bibsource    = {dblp computer science bibliography, https://dblp.org}
}

@inproceedings{DBLP:conf/naacl/YanAV25,
  author       = {Minghao Yan and
                  Saurabh Agarwal and
                  Shivaram Venkataraman},
  editor       = {Luis Chiruzzo and
                  Alan Ritter and
                  Lu Wang},
  title        = {Decoding Speculative Decoding},
  booktitle    = {Proceedings of the 2025 Conference of the Nations of the Americas
                  Chapter of the Association for Computational Linguistics: Human Language
                  Technologies, {NAACL} 2025 - Volume 1: Long Papers, Albuquerque, New
                  Mexico, USA, April 29 - May 4, 2025},
  pages        = {6460--6473},
  publisher    = {Association for Computational Linguistics},
  year         = {2025},
  url          = {https://doi.org/10.18653/v1/2025.naacl-long.328},
  doi          = {10.18653/V1/2025.NAACL-LONG.328},
  bibsource    = {dblp computer science bibliography, https://dblp.org}
}

@inproceedings{DBLP:conf/asplos/XuPWZY0026,
  author       = {Jiaming Xu and
                  Jiayi Pan and
                  Hanzhen Wang and
                  Yongkang Zhou and
                  Jiancai Ye and
                  Yu Wang and
                  Guohao Dai},
  editor       = {Benjamin C. Lee and
                  Harry Xu and
                  Mark Silberstein and
                  Bingyao Li},
  title        = {\emph{SpeContext: } Enabling Efficient Long-context Reasoning with
                  Speculative Context Sparsity in LLMs},
  booktitle    = {Proceedings of the 31st {ACM} International Conference on Architectural
                  Support for Programming Languages and Operating Systems, Volume 2,
                  {ASPLOS} 2026, Pittsburgh, PA, USA, March 22-26, 2026},
  pages        = {1832--1847},
  publisher    = {{ACM}},
  year         = {2026},
  url          = {https://doi.org/10.1145/3779212.3790224},
  doi          = {10.1145/3779212.3790224},
  bibsource    = {dblp computer science bibliography, https://dblp.org}
}

@article{DBLP:journals/corr/abs-2505-22101,
  author       = {Zhiyu Li and
                  Shichao Song and
                  Hanyu Wang and
                  Simin Niu and
                  Ding Chen and
                  Jiawei Yang and
                  Chenyang Xi and
                  Huayi Lai and
                  Jihao Zhao and
                  Yezhaohui Wang and
                  Junpeng Ren and
                  Zehao Lin and
                  Jiahao Huo and
                  Tianyi Chen and
                  Kai Chen and
                  Kehang Li and
                  Zhiqiang Yin and
                  Qingchen Yu and
                  Bo Tang and
                  Hongkang Yang and
                  Zhi{-}Qin John Xu and
                  Feiyu Xiong},
  title        = {MemOS: An Operating System for Memory-Augmented Generation {(MAG)}
                  in Large Language Models},
  journal      = {CoRR},
  volume       = {abs/2505.22101},
  year         = {2025},
  url          = {https://doi.org/10.48550/arXiv.2505.22101},
  doi          = {10.48550/ARXIV.2505.22101},
  eprinttype    = {arXiv},
  eprint       = {2505.22101},
  bibsource    = {dblp computer science bibliography, https://dblp.org}
}

@inproceedings{DBLP:conf/iclr/FuCAX0X25,
  author       = {Yu Fu and
                  Zefan Cai and
                  Abedelkadir Asi and
                  Wayne Xiong and
                  Yue Dong and
                  Wen Xiao},
  title        = {Not All Heads Matter: {A} Head-Level {KV} Cache Compression Method
                  with Integrated Retrieval and Reasoning},
  booktitle    = {The Thirteenth International Conference on Learning Representations,
                  {ICLR} 2025, Singapore, April 24-28, 2025},
  publisher    = {OpenReview.net},
  year         = {2025},
  url          = {https://openreview.net/forum?id=FJFVmeXusW},
  bibsource    = {dblp computer science bibliography, https://dblp.org}
}

@inproceedings{DBLP:conf/sosp/ZhangHZLC25,
  author       = {Yanqi Zhang and
                  Yuwei Hu and
                  Runyuan Zhao and
                  John C. S. Lui and
                  Haibo Chen},
  editor       = {Youjip Won and
                  Youngjin Kwon and
                  Ding Yuan and
                  Rebecca Isaacs},
  title        = {DiffKV: Differentiated Memory Management for Large Language Models
                  with Parallel {KV} Compaction},
  booktitle    = {Proceedings of the {ACM} {SIGOPS} 31st Symposium on Operating Systems
                  Principles, {SOSP} 2025, Lotte Hotel World, Seoul, Republic of Korea,
                  October 13-16, 2025},
  pages        = {431--445},
  publisher    = {{ACM}},
  year         = {2025},
  url          = {https://doi.org/10.1145/3731569.3764810},
  doi          = {10.1145/3731569.3764810},
  bibsource    = {dblp computer science bibliography, https://dblp.org}
}

@inproceedings{DBLP:conf/nips/LiHYVLYCLC24,
  author       = {Yuhong Li and
                  Yingbing Huang and
                  Bowen Yang and
                  Bharat Venkitesh and
                  Acyr Locatelli and
                  Hanchen Ye and
                  Tianle Cai and
                  Patrick Lewis and
                  Deming Chen},
  editor       = {Amir Globersons and
                  Lester Mackey and
                  Danielle Belgrave and
                  Angela Fan and
                  Ulrich Paquet and
                  Jakub M. Tomczak and
                  Cheng Zhang},
  title        = {SnapKV: {LLM} Knows What You are Looking for Before Generation},
  booktitle    = {Advances in Neural Information Processing Systems 37: Annual Conference
                  on Neural Information Processing Systems 2024, NeurIPS 2024, Vancouver,
                  BC, Canada, December 10 - 15, 2024},
  year         = {2024},
  url          = {http://papers.nips.cc/paper\_files/paper/2024/hash/28ab418242603e0f7323e54185d19bde-Abstract-Conference.html},
  bibsource    = {dblp computer science bibliography, https://dblp.org}
}

@inproceedings{DBLP:conf/acl/YangHGHZ024,
  author       = {Dongjie Yang and
                  Xiaodong Han and
                  Yan Gao and
                  Yao Hu and
                  Shilin Zhang and
                  Hai Zhao},
  editor       = {Lun{-}Wei Ku and
                  Andre Martins and
                  Vivek Srikumar},
  title        = {PyramidInfer: Pyramid {KV} Cache Compression for High-throughput {LLM}
                  Inference},
  booktitle    = {Findings of the Association for Computational Linguistics, {ACL} 2024,
                  Bangkok, Thailand and virtual meeting, August 11-16, 2024},
  series       = {Findings of {ACL}},
  pages        = {3258--3270},
  publisher    = {Association for Computational Linguistics},
  year         = {2024},
  url          = {https://doi.org/10.18653/v1/2024.findings-acl.195},
  doi          = {10.18653/V1/2024.FINDINGS-ACL.195},
  bibsource    = {dblp computer science bibliography, https://dblp.org}
}

@article{grattafiori2024llama3herdmodels,
       author = {Llama Team, {Meta AI}},
        title = "{The Llama 3 Herd of Models}",
      journal = {arXiv e-prints},
         year = 2024,
        month = jul,
          eid = {arXiv:2407.21783},
        pages = {arXiv:2407.21783},
          doi = {10.48550/arXiv.2407.21783},
archivePrefix = {arXiv},
       eprint = {2407.21783},
 primaryClass = {cs.AI},
       adsurl = {https://ui.adsabs.harvard.edu/abs/2024arXiv240721783D}
}

@misc{qwen2025qwen25technicalreport,
      title={Qwen2.5 Technical Report}, 
      author={Qwen and : and An Yang and Baosong Yang and Beichen Zhang and Binyuan Hui and Bo Zheng and Bowen Yu and Chengyuan Li and Dayiheng Liu and Fei Huang and Haoran Wei and Huan Lin and Jian Yang and Jianhong Tu and Jianwei Zhang and Jianxin Yang and Jiaxi Yang and Jingren Zhou and Junyang Lin and Kai Dang and Keming Lu and Keqin Bao and Kexin Yang and Le Yu and Mei Li and Mingfeng Xue and Pei Zhang and Qin Zhu and Rui Men and Runji Lin and Tianhao Li and Tianyi Tang and Tingyu Xia and Xingzhang Ren and Xuancheng Ren and Yang Fan and Yang Su and Yichang Zhang and Yu Wan and Yuqiong Liu and Zeyu Cui and Zhenru Zhang and Zihan Qiu},
      year={2025},
      eprint={2412.15115},
      archivePrefix={arXiv},
      primaryClass={cs.CL},
      url={https://arxiv.org/abs/2412.15115}, 
}

@misc{hsieh2024rulerwhatsrealcontext,
      title={RULER: What's the Real Context Size of Your Long-Context Language Models?}, 
      author={Cheng-Ping Hsieh and Simeng Sun and Samuel Kriman and Shantanu Acharya and Dima Rekesh and Fei Jia and Yang Zhang and Boris Ginsburg},
      year={2024},
      eprint={2404.06654},
      archivePrefix={arXiv},
      primaryClass={cs.CL},
      url={https://arxiv.org/abs/2404.06654}, 
}

@article{DBLP:journals/corr/abs-2409-10516,
  author       = {Di Liu and
                  Meng Chen and
                  Baotong Lu and
                  Huiqiang Jiang and
                  Zhenhua Han and
                  Qianxi Zhang and
                  Qi Chen and
                  Chengruidong Zhang and
                  Bailu Ding and
                  Kai Zhang and
                  Chen Chen and
                  Fan Yang and
                  Yuqing Yang and
                  Lili Qiu},
  title        = {RetrievalAttention: Accelerating Long-Context {LLM} Inference via
                  Vector Retrieval},
  journal      = {CoRR},
  volume       = {abs/2409.10516},
  year         = {2024},
  url          = {https://doi.org/10.48550/arXiv.2409.10516},
  doi          = {10.48550/ARXIV.2409.10516},
  eprinttype   = {arXiv},
  eprint       = {2409.10516},
  bibsource    = {dblp computer science bibliography, https://dblp.org}
}

@inproceedings{DBLP:conf/icml/RibarCHBLO24,
  author       = {Luka Ribar and
                  Ivan Chelombiev and
                  Luke Hudlass{-}Galley and
                  Charlie Blake and
                  Carlo Luschi and
                  Douglas Orr},
  editor       = {Ruslan Salakhutdinov and
                  Zico Kolter and
                  Katherine A. Heller and
                  Adrian Weller and
                  Nuria Oliver and
                  Jonathan Scarlett and
                  Felix Berkenkamp},
  title        = {SparQ Attention: Bandwidth-Efficient {LLM} Inference},
  booktitle    = {Forty-first International Conference on Machine Learning, {ICML} 2024,
                  Vienna, Austria, July 21-27, 2024},
  series       = {Proceedings of Machine Learning Research},
  pages        = {42558--42583},
  publisher    = {{PMLR} / OpenReview.net},
  year         = {2024},
  url          = {https://proceedings.mlr.press/v235/ribar24a.html},
  bibsource    = {dblp computer science bibliography, https://dblp.org}
}

@inproceedings{DBLP:conf/osdi/LeeLSS24,
  author       = {Wonbeom Lee and
                  Jungi Lee and
                  Junghwan Seo and
                  Jaewoong Sim},
  editor       = {Ada Gavrilovska and
                  Douglas B. Terry},
  title        = {InfiniGen: Efficient Generative Inference of Large Language Models
                  with Dynamic {KV} Cache Management},
  booktitle    = {18th {USENIX} Symposium on Operating Systems Design and Implementation,
                  {OSDI} 2024, Santa Clara, CA, USA, July 10-12, 2024},
  pages        = {155--172},
  publisher    = {{USENIX} Association},
  year         = {2024},
  url          = {https://www.usenix.org/conference/osdi24/presentation/lee},
  bibsource    = {dblp computer science bibliography, https://dblp.org}
}

@inproceedings{DBLP:conf/icml/LiuYJZXBC024,
  author       = {Zirui Liu and
                  Jiayi Yuan and
                  Hongye Jin and
                  Shaochen (Henry) Zhong and
                  Zhaozhuo Xu and
                  Vladimir Braverman and
                  Beidi Chen and
                  Xia Hu},
  editor       = {Ruslan Salakhutdinov and
                  Zico Kolter and
                  Katherine A. Heller and
                  Adrian Weller and
                  Nuria Oliver and
                  Jonathan Scarlett and
                  Felix Berkenkamp},
  title        = {{KIVI:} {A} Tuning-Free Asymmetric 2bit Quantization for {KV} Cache},
  booktitle    = {Forty-first International Conference on Machine Learning, {ICML} 2024,
                  Vienna, Austria, July 21-27, 2024},
  series       = {Proceedings of Machine Learning Research},
  pages        = {32332--32344},
  publisher    = {{PMLR} / OpenReview.net},
  year         = {2024},
  url          = {https://proceedings.mlr.press/v235/liu24bz.html},
  bibsource    = {dblp computer science bibliography, https://dblp.org}
}

@inproceedings{DBLP:conf/icml/TangZZXKH24,
  author       = {Jiaming Tang and
                  Yilong Zhao and
                  Kan Zhu and
                  Guangxuan Xiao and
                  Baris Kasikci and
                  Song Han},
  title        = {{QUEST:} Query-Aware Sparsity for Efficient Long-Context {LLM} Inference},
  booktitle    = {Forty-first International Conference on Machine Learning, {ICML} 2024,
                  Vienna, Austria, July 21-27, 2024},
  publisher    = {OpenReview.net},
  year         = {2024},
  url          = {https://openreview.net/forum?id=KzACYw0MTV},
  bibsource    = {dblp computer science bibliography, https://dblp.org}
}


\appendix
\begin{sloppypar}
\section*{Appendix}

\appendix

\section{\system shows comparable serving latencies.}
\label{sec: serving latencies}

\begin{figure}[H]
    \centering
    \subfigure[TTFT comparison.]{
        \label{fig: eval/ttft}
        \hspace{-0.28cm}
        \includegraphics[width=0.48\linewidth]{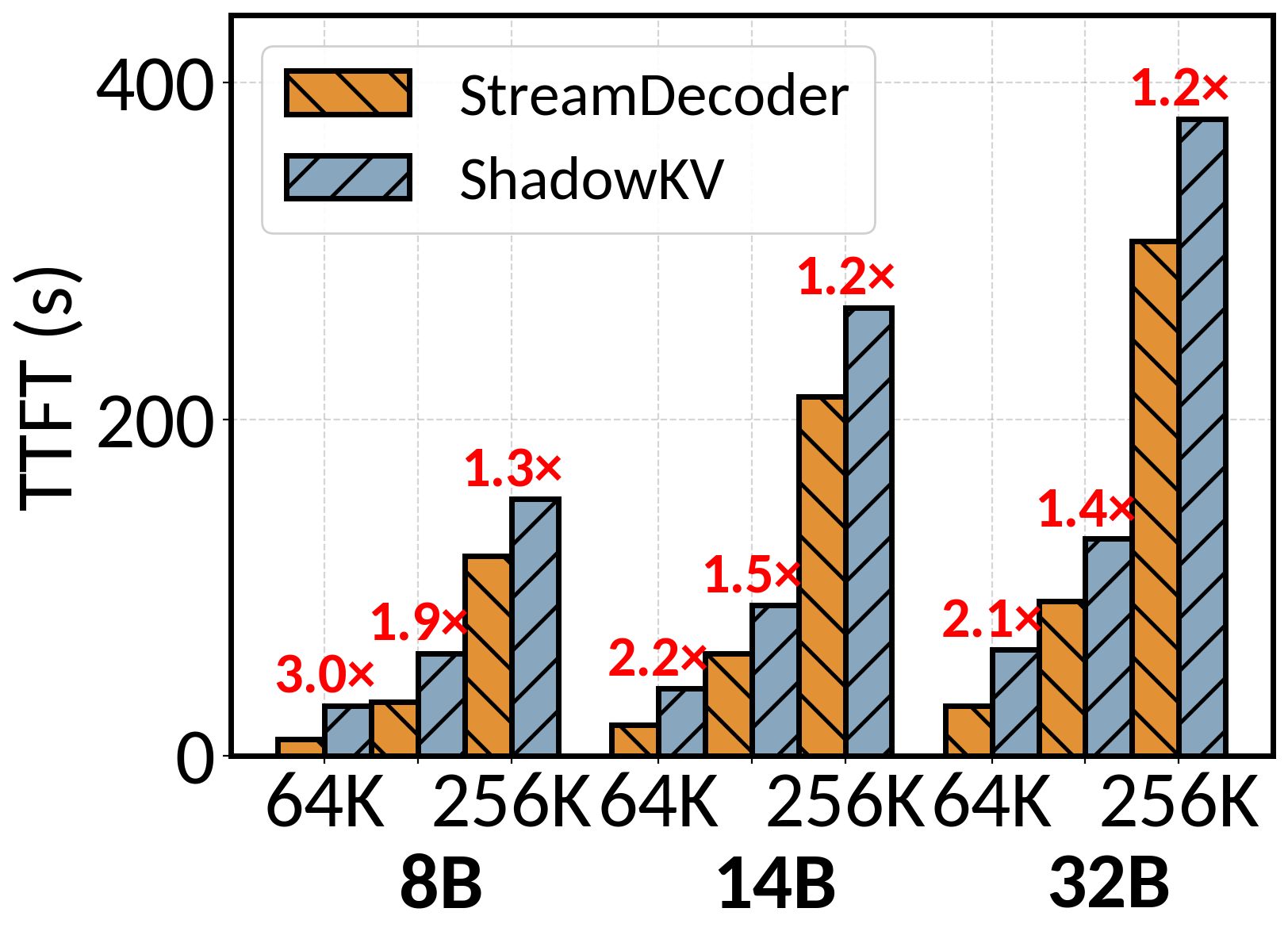}
    }
   \hspace{-0.2cm}
    \subfigure[TPOT comparison.]{
        \label{fig: eval/tpot}
        \includegraphics[width=0.48\linewidth]{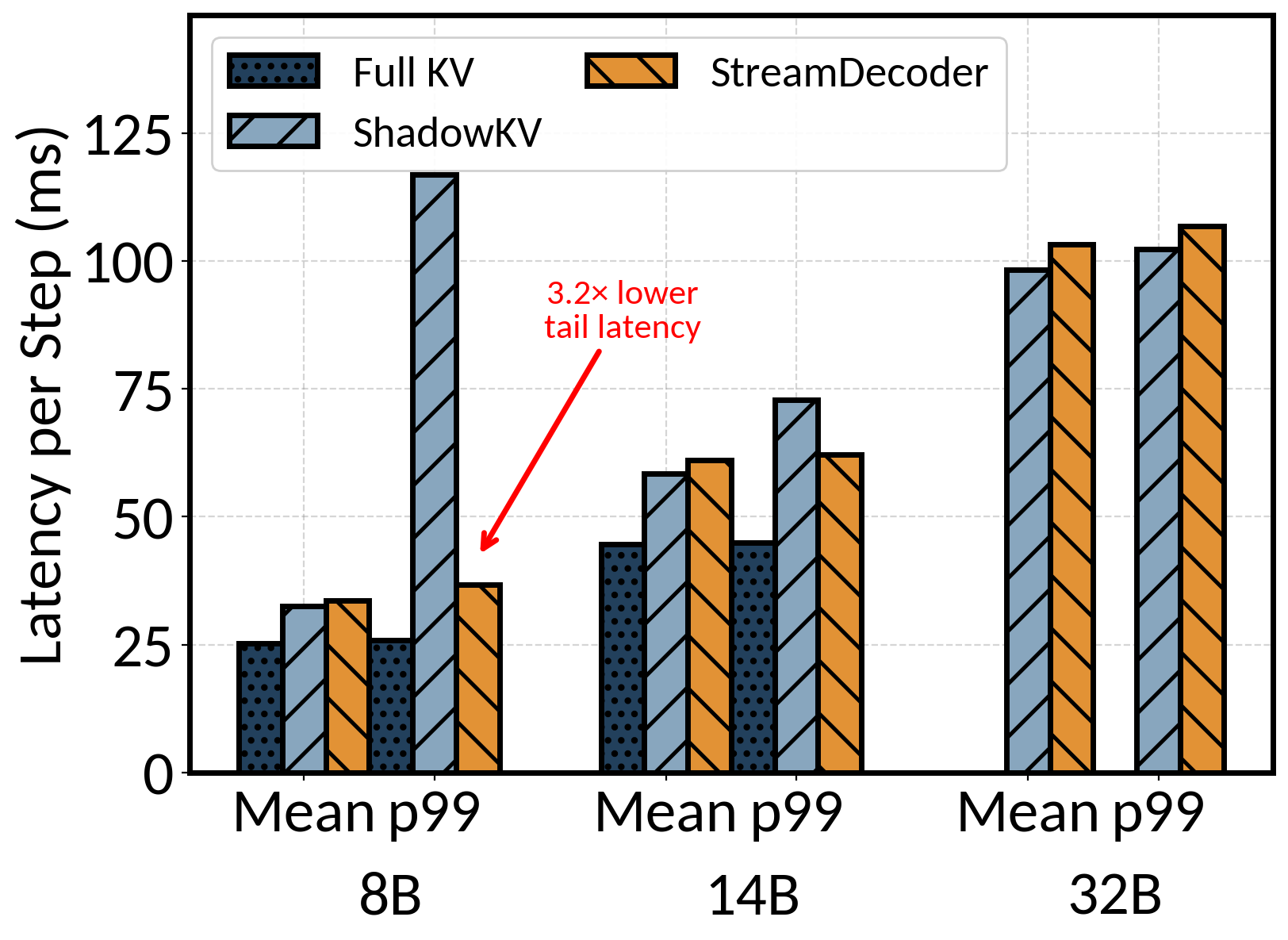}
    }
    \vspace{-0.3cm}
    \label{fig: eval/latency compare}
    \caption{TTFT and TPOT comparison across different systems.}
    \vspace{-0.4cm}
\end{figure}

Unlike traditional dense inference baselines (such as vLLM) that scale poorly under long-context constraints, or sparse KV cache baselines (such as ShadowKV and SpeCache) that suffer from significant runtime memory stalls or reconstruction overheads on the critical path, StreamDecoder utilizes a dual-token decoding pipeline alongside a layer-aware predictive prefetching schedule to balance both phases efficiently. As shown in Figure\ref{fig: eval/ttft},\ref{fig: eval/tpot} Our system reduces over \adddata{$3\times$} of TTFT reduction and have the almost same TPOT compared with the state-of-the-art systems.

\label{totalpage}
\end{sloppypar}

\end{document}